\documentclass[12pt]{article}
\pdfoutput=1
\usepackage{jheppub}
\usepackage{epsfig}
\usepackage{amsmath}
\usepackage{amssymb}
\usepackage{amsfonts}
\usepackage{amsxtra}
\usepackage{amsthm}
\usepackage{mathrsfs}
\usepackage{makeidx}
\usepackage{graphics}
\usepackage{dsfont}
\usepackage{mathtools}
\usepackage{graphicx}
\usepackage{subcaption}
\usepackage{placeins}
\usepackage{bm}
\usepackage[capitalise]{cleveref}
\usepackage{empheq}
\usepackage{colortbl}
\usepackage{xcolor}
\usepackage{enumerate}
\usepackage{titlesec}
\usepackage{longtable}
\usepackage{float}
\usepackage{color}
\usepackage{tikz}
\usepackage{xfrac}
\usepackage{footnote}
\usepackage{rotating}
\usepackage{lscape}
\usepackage{makecell}
\usepackage{environ}
\usepackage{tabularx}
\usepackage{subfiles}
\usepackage[export]{adjustbox}
\usepackage{ytableau}
\usepackage{tikz-3dplot}
\usepackage{slashed}
\usepackage{pifont}
\usepackage{multirow}
\usepackage{mdframed}
\usepackage{bbm}
\usepackage[normalem]{ulem}
\usepackage{enumitem}

\usetikzlibrary{positioning,trees,decorations.pathmorphing,decorations.markings,decorations.pathreplacing,calc,shapes,patterns,arrows,chains,arrows.meta,fit,fadings,decorations.markings,graphs,graphs.standard,quotes}

\titleformat{\paragraph}
{\normalfont\normalsize\bfseries}{\theparagraph}{1em}{}
\titlespacing*{\paragraph}{0pt}{3.25ex plus 1ex minus .2ex}{1.5ex plus .2ex}

\DeclareMathOperator{\U}{U}
\DeclareMathOperator{\SU}{SU}
\DeclareMathOperator{\SO}{SO}

\DeclareMathOperator{\USp}{USp}

\newcommand{\coma}{\, , \quad}
\newcommand{\fstop}{\, .}

\newcommand{\aand}{\quad\text{and}\quad}

\def\RR{{\mathbb{R}}}
\def\ZZ{{\mathbb{Z}}}

\theoremstyle{definition}

\catcode`\@=11   
\newdimen\@rotdimen
\newbox\@rotbox  

\def\@vspec#1{\special{ps:#1}}
\def\@rotstart#1{\@vspec{gsave currentpoint currentpoint translate
   #1 neg exch neg exch translate}}
\def\@rotfinish{\@vspec{currentpoint grestore moveto}}
\def\@rotr#1{\@rotdimen=\ht#1\advance\@rotdimen by\dp#1%
   \hbox to\@rotdimen{\hskip\ht#1\vbox to\wd#1{\@rotstart{90 rotate}%
   \box#1\vss}\hss}\@rotfinish}
\def\@rotl#1{\@rotdimen=\ht#1\advance\@rotdimen by\dp#1%
   \hbox to\@rotdimen{\vbox to\wd#1{\vskip\wd#1\@rotstart{270 rotate}%
   \box#1\vss}\hss}\@rotfinish}%
\def\@rotu#1{\@rotdimen=\ht#1\advance\@rotdimen by\dp#1%
   \hbox to\wd#1{\hskip\wd#1\vbox to\@rotdimen{\vskip\@rotdimen
   \@rotstart{-1 dup scale}\box#1\vss}\hss}\@rotfinish}%
\def\@rotf#1{\hbox to\wd#1{\hskip\wd#1\@rotstart{-1 1 scale}%
   \box#1\hss}\@rotfinish}%
\def\rotate{\@ifnextchar[{\@rotate}{\@rotate[l\right]}}
\def\@rotate[#1]#2{\setbox\@rotbox=\hbox{#2}\@nameuse{@rot#1}\@rotbox}

\catcode`\@=12

\usetikzlibrary{positioning}
\usetikzlibrary{chains}
\usetikzlibrary{arrows, arrows.meta ,fit,decorations.pathreplacing}
\tikzstyle{every picture}+=[remember picture]
\tikzstyle{na} = [baseline]
\tikzstyle{ligne}=[draw, thick]

\usetikzlibrary{arrows, decorations.markings, calc, fadings, decorations.pathreplacing, patterns, decorations.pathmorphing, positioning}
\tikzset{>={Latex[width=1.5mm,length=1.5mm]}}
\tikzset{bd/.style={circle, draw=black, inner sep=0pt, fill=black, minimum size=1.2mm}}
\tikzset{bld/.style={circle, draw=blue, inner sep=0pt, fill=blue, minimum size=1.2mm}}
\tikzset{wd/.style={circle, draw=black, inner sep=0pt, fill=white, minimum size=1.2mm}}
\tikzset{rd/.style={circle, draw=red, inner sep=0pt, fill=red, minimum size=.9mm}}
\tikzset{wrd/.style={circle, draw=red, inner sep=0pt, fill=white, minimum size=.9mm}}
\usetikzlibrary{graphs,graphs.standard,quotes}

\def\node#1#2{\overset{#1}{\underset{#2}{{\color{gray} \bullet}}}}

\def\node#1#2{\overset{#1}{\underset{#2}{\circ}}}

\tikzstyle{every picture}+=[remember picture]
\tikzstyle{na} = [baseline=-.5ex]

\newcommand{\ie}{i.e.}

\numberwithin{equation}{section}
\newcommand{\bes}[1]{\begin{equation} \begin{split} #1\end{split} \end{equation}}

\newcommand{\nn}{\nonumber}

\newcommand{\be}{\begin{equation}} \newcommand{\ee}{\end{equation}}
\newcommand{\bea}{\begin{equation} \begin{aligned}} \newcommand{\eea}{\end{aligned} \end{equation}}

\def\tilde{\widetilde}

\def\bar{\overline}

\def\rt2{\sqrt{2}}

\def\mod{{\rm mod}}

\def\CN{{\cal N}}

\def\CS{{\cal S}}

\def\1{{\ds 1}}

\newcommand{\fm}{\mathfrak{m}}
\newcommand{\fn}{\mathfrak{n}}

\def\SO{\mathrm{SO}}

\def\SU{\mathrm{SU}}

\def\fN{\mathfrak{N}}

\def\repa{\raise4pt\hbox{$\square$}\mkern-14mu\raise-4pt\hbox{$\square$}}
\def\repab{\overline{\raise4pt\hbox{$\square$}\mkern-14mu\raise-4pt\hbox{$\square$}\mkern-1mu}}

\def\smileface{\ensuremath{\hbox{\large$\bigcirc$}\mkern-15mu\raise-1pt\hbox{\scriptsize$\smallsmile$}%
\mkern-10mu\raise4pt\hbox{..}\mkern4mu}}
\def\frownface{\ensuremath{\hbox{\large$\bigcirc$}\mkern-15mu\raise-1pt\hbox{\scriptsize$\smallfrown$}%
\mkern-10mu\raise4pt\hbox{..}\mkern4mu}}

\newcommand{\ba}{\begin{array}}
\newcommand{\ea}{\end{array}}
\newcommand{\bi}{\begin{itemize}}
\newcommand{\ei}{\end{itemize}}

\def\bea#1\eea{\allowdisplaybreaks \begin{align}#1\end{align}}
 \newcommand{\ben}{\begin{enumerate}}
\newcommand{\een}{\end{enumerate}}
\newcommand{\bean}{\begin{eqnarray*}}
\newcommand{\eean}{\end{eqnarray*}}
\newcommand{\eref}[1]{(\ref{#1})}

\newcommand{\BZ}{\mathbb{Z}}

\newcommand{\BH}{\mathbb{H}}

\definecolor{light-gray}{gray}{0.5}

\newcommand{\blue}{\color{blue}}
\newcommand{\gray}{\color{light-gray}}
\newcommand{\red}{\color{red}}

\def\aup#1 {\overset{#1}{\uparrow} \, \overset{\tilde{#1}}{\downarrow}}

\tikzset{snake it/.style={decorate, decoration={snake, amplitude=.4mm, segment length=2mm,
                       post length=0mm,pre length=0mm}}}
                       
 \newcommand{\GCD}{\mathrm{GCD}}

\hypersetup{
	pdftitle={Conformal Manifolds and 3d Mirrors of $(D_n,D_m)$ Theories},    
	pdfauthor={\textcopyright\ Federico Carta, Simone Giacomelli, Noppadol Mekareeya, Alessandro Mininno},     
	pdfsubject={HEP},   
	pdfcreator={pdfLaTex},   
	pdfproducer={LaTex}, 
	pdfkeywords={},
	colorlinks=true,
}

\makeatletter
\newsavebox{\measure@tikzpicture}
\NewEnviron{scaletikzpicturetowidth}[1]{%
  \def\tikz@width{#1}%
  \begin{lrbox}{\measure@tikzpicture}%
  \BODY
  \end{lrbox}%
  \pgfmathparse{#1/\wd\measure@tikzpicture}%
  \BODY
}
\makeatother

\makeatletter
\def\squarecorner#1{
    \pgf@x=\the\wd\pgfnodeparttextbox%
    \pgfmathsetlength\pgf@xc{\pgfkeysvalueof{/pgf/inner xsep}}%
    \advance\pgf@x by 2\pgf@xc%
    \pgfmathsetlength\pgf@xb{\pgfkeysvalueof{/pgf/minimum width}}%
    \ifdim\pgf@x<\pgf@xb%
        \pgf@x=\pgf@xb%
    \fi%
    \pgf@y=\ht\pgfnodeparttextbox%
    \advance\pgf@y by\dp\pgfnodeparttextbox%
    \pgfmathsetlength\pgf@yc{\pgfkeysvalueof{/pgf/inner ysep}}%
    \advance\pgf@y by 2\pgf@yc%
    \pgfmathsetlength\pgf@yb{\pgfkeysvalueof{/pgf/minimum height}}%
    \ifdim\pgf@y<\pgf@yb%
        \pgf@y=\pgf@yb%
    \fi%
    \ifdim\pgf@x<\pgf@y%
        \pgf@x=\pgf@y%
    \else
        \pgf@y=\pgf@x%
    \fi
    \pgf@x=#1.5\pgf@x%
    \advance\pgf@x by.5\wd\pgfnodeparttextbox%
    \pgfmathsetlength\pgf@xa{\pgfkeysvalueof{/pgf/outer xsep}}%
    \advance\pgf@x by#1\pgf@xa%
    \pgf@y=#1.5\pgf@y%
    \advance\pgf@y by-.5\dp\pgfnodeparttextbox%
    \advance\pgf@y by.5\ht\pgfnodeparttextbox%
    \pgfmathsetlength\pgf@ya{\pgfkeysvalueof{/pgf/outer ysep}}%
    \advance\pgf@y by#1\pgf@ya%
}
\makeatother

\pgfdeclareshape{square}{
    \savedanchor\northeast{\squarecorner{}}
    \savedanchor\southwest{\squarecorner{-}}

    \foreach \x in {east,west} \foreach \y in {north,mid,base,south} {
        \inheritanchor[from=rectangle]{\y\space\x}
    }
    \foreach \x in {east,west,north,mid,base,south,center,text} {
        \inheritanchor[from=rectangle]{\x}
    }
    \inheritanchorborder[from=rectangle]
    \inheritbackgroundpath[from=rectangle]
}

\tikzset{stretch/.initial=1}
\newcommand\drawloop[4][]%
   {\draw[shorten <=0pt, shorten >=0pt,#1]
      ($(#2)!\pgfkeysvalueof{/tikz/stretch}!(#2.#3)$)
      let \p1=($(#2.center)!\pgfkeysvalueof{/tikz/stretch}!(#2.north)-(#2)$),
          \n1= {veclen(\x1,\y1)*sin(0.5*(#4-#3))/sin(0.5*(180-#4+#3))}
      in arc [start angle={#3-90}, end angle={#4+90}, radius=\n1]%
   }

\preprint{IFT-UAM/CSIC-21-109, ZMP-HH/21-20}
\title{Conformal Manifolds and 3d Mirrors of $(D_n,D_m)$ Theories}
\author[a]{Federico Carta,}
\author[b,c]{~Simone Giacomelli,}
\author[c,d,e]{~Noppadol Mekareeya}
\author[f,g]{\\ and Alessandro Mininno}
\affiliation[a]{Department of Mathematical Sciences,
		Durham University, \\ Durham, DH1 3LE, United Kingdom}
\affiliation[b]{Mathematical Institute, University of Oxford, Woodstock Road, Oxford, \\ OX2 6GG, United Kingdom}
\affiliation[c]{Dipartimento di Fisica, Universit\`a di Milano-Bicocca, Piazza della Scienza 3, \\ I-20126 Milano, Italy}
\affiliation[d]{INFN, sezione di Milano-Bicocca, Piazza della Scienza 3, \\ I-20126 Milano, Italy}
\affiliation[e]{Department of Physics, Faculty of Science, Chulalongkorn University, \\ Phayathai Road,
Pathumwan, Bangkok 10330, Thailand}
\affiliation[f]{Instituto de F\'{\i}sica Te\'orica IFT-UAM/CSIC,\\
		C/ Nicol\'as Cabrera 13-15, 
		Campus de Cantoblanco, 28049 Madrid, Spain}
\affiliation[g]{II. Institut f\"ur Theoretische Physik, Universit\"at Hamburg,\\
Luruper Chaussee 149, 22607 Hamburg, Germany}
\emailAdd{federico.carta@durham.ac.uk}
\emailAdd{simone.giacomelli@unimib.it}
\emailAdd{n.mekareeya@gmail.com}
\emailAdd{alessandro.mininno@desy.de}
\abstract{The Argyres-Douglas (AD) theories of type $(D_n,D_m)$, realized by type IIB geometrical engineering on a single hypersurface singularity, are studied. We analyze their conformal manifolds and propose the 3d mirror theories of all theories in this class upon reduction on a circle. A subclass of the AD theories in question that admits marginal couplings is found to be $\mathrm{SO}$ or $\mathrm{USp}$ gaugings of certain $D_p(\mathrm{SO}(2N))$ and $D_p(\mathrm{USp}(2N))$ theories. For such theories, we develop a method to derive this weakly-coupled description from the Newton polygon associated to the singularity. We further find that the presence of crepant resolutions of the geometry is reflected in the presence of a (non-abelian) symplectic-type gauge node in the quiver description of the 3d mirror theory. The other important results include the 3d mirrors of all $D_p(\mathrm{SO}(2N))$ theories, as well as certain properties of the $D_p(\mathrm{USp}(2N))$ theories that admit Lagrangian descriptions.}

\begin{document} 

\maketitle

\section{Introduction}

Since the discovery of the Seiberg-Witten (SW) solution \cite{Seiberg:1994rs, Seiberg:1994aj}, four-dimensional $\mathcal{N}=2$ theories have attracted a lot of attention. This is often due to the fact that the large amount of supersymmetry constrains the dynamics enough to make these models a useful theoretical laboratory for the exploration of nonperturbative dynamics. Remarkably, the physics on the Coulomb branch (CB) of these theories is accessible even for nonlagrangian models and this fact led to the discovery of a large class of intrinsically strongly-coupled theories, first as low-energy theories at singular points of the Coulomb branch of $\mathcal{N}=2$ gauge theories \cite{Argyres:1995jj, Argyres:1995xn, Eguchi:1996vu, Eguchi:1996ds} and then more abstractly via geometric methods, either as compactifications of higher dimensional superconformal theories (SCFTs) \cite{Gaiotto:2009we, Gaiotto:2009hg, Ohmori:2015pua, Ohmori:2015pia, Ohmori:2018ona} or from the compactification of superstring theory on local Calabi-Yau (CY) $3$-folds \cite{Katz:1996fh, Shapere:1999xr}. In this latter case the geometric engineering in Type IIB is particularly convenient since the complex structure moduli of the geometry (i.e. classical properties) encode the information about the quantum-corrected Coulomb branch physics.  

A natural question is then how the stringy geometry encodes the information about the Higgs branch (HB) of the four-dimensional theory. This is harder to address since a classical Type IIB analysis is not enough to provide the answer. An effective strategy to make progress in this direction is to describe the Higgs branch of the 4d theory as the Coulomb branch of a three-dimensional theory with eight supercharges, the so-called 3d mirror dual \cite{Intriligator:1996ex} or magnetic quiver \cite{Ferlito:2017xdq, Cabrera:2018jxt,  Hanany:2018uhm, Cabrera:2019izd, Akhond:2020vhc, Akhond:2021knl, Akhond:2021ffo, Bourget:2020gzi, Bourget:2020asf, Bourget:2020mez} in more modern terminology. Following previous work (see e.g. \cite{boalch2008irregular, Xie:2012hs, Xie:2013jc, DelZotto:2014kka, Buican:2015hsa, Xie:2017vaf, Benvenuti:2017kud, Benvenuti:2017bpg, Dey:2020hfe, Closset:2020scj, Closset:2020afy, Giacomelli:2020ryy, Carta:2021whq, Xie:2021ewm, Dey:2021rxw}) our goal is to construct the 3d mirror theory in the case of local Calabi-Yau geometries described by hypersurface singularities in $\mathbb{C}^4$. More specifically, in this note we consider hypersurfaces given by the sum of two ADE singularities (usually referred to as $(G,G')$ models \cite{Cecotti:2010fi}), where both singularities are of type $D$. 

The analysis of the present work represents a natural continuation of \cite{Giacomelli:2020ryy, Carta:2021whq}, where $(A,A)$ and $(A,D)$ theories have been considered, and with respect to the cases already discussed in the literature presents a new hurdle: The $(D,D)$ singularities are not terminal and we have crepant resolutions, which imply a mismatch between the dimension of the Higgs branch and the number of mass parameters since crepant divisors contribute to the former but not to the latter (see the discussion in the Introduction of \cite{Closset:2020scj}). As a result, the 3d mirror cannot simply be an abelian gauge theory as it was in the $(A,A)$ and $(A,D)$ cases. This would indeed immediately imply that the Higgs branch dimension and the rank of the global symmetry of the 4d theory agree. We find that the difference is accounted for by the presence in the 3d mirror of a balanced $\USp(2n)$ gauge node, whose topological symmetry \cite{Kapustin:1998fa, Gaiotto:2008ak} contributes one to the rank of the global symmetry of the 4d theory and whose rank contributes $n$ to the dimension of the Higgs branch of the 4d theory. This represents a new conceptual step towards a systematic understanding of the Higgs branch of 4d SCFTs from hypersurface singularities. 

The strategy we apply in this work to extract the 3d mirrors of $(D_n,D_m)$ theories is to start from the already understood $(A,A)$ and $(A,D)$ cases and to construct the 3d quivers step by step for a large set of examples. This then allows us to guess the general answer. In order to implement this program, we first need to study the conformal manifold of these theories and identify weakly-coupled cusps (see also \cite{Buican:2014hfa, Buican:2017fiq, Buican:2021xhs}), which turn out to involve gaugings of $D_p(\SO)$ or $D_p(\USp)$ theories introduced in \cite{Cecotti:2012jx, Cecotti:2013lda, Wang:2015mra, Wang:2018gvb} (see also~\cite{Carta:2021whq}). This therefore reduces the problem to gauging building blocks whose 3d mirror is already known. The key feature for identifying the structure of the conformal manifold is the fact that from the geometry we can easily extract the Seiberg-Witten curve of the theory, at least in a certain limit. This is due to the specific structure of the family of hypersurface singularities which engineers the $(D_n,D_m)$ theories, which allows us to plot on a plane the various deformations of the theory. 

Let us describe salient features of the 3d mirrors of $(D_n, D_m)$ theories. For generic values of $n$ and $m$, such a theory consists of a collection of free hypermultiplets, along with an interacting 3d $\CN=4$ SCFT that admits a quiver description. We first discuss the latter. The quiver consists of a balanced central node of the $\USp$-type, which is surrounded by a collection of $\SO(2)\cong D_1$ gauge nodes, possibly with certain number of flavors of hypermultiplets.  A subset of such $D_1$ gauge nodes are connected together by lines to form a complete graph. The central $\USp$ gauge node is connected to the surrounding $D_1$ nodes in a highly non-trivial way, described in detail in the main text.  We now describe the origin and properties of such free hypermultiplets.  As described in the context of the $(A_n, A_m)$ and $(A_n, D_m)$ theories \cite{Nanopoulos:2010bv, Giacomelli:2020ryy, Carta:2021whq}, the free sector arises from dimensional reduction of the SCFTs with no Higgs branch, also known as the non-Higgsable SCFTs, that are present at a generic point of the Higgs branch of the 4d theory.  The readers are referred to \cite[Appendix C]{Carta:2021whq} for an extensive list of the non-Higgsable SCFTs and their properties. In particular, the difference between the rank of the 4d theory and the Higgs branch of the aforementioned quiver is equal to the number of the free hypermultiplets and thus the rank of the non-Higgsable SCFTs in question. Moreover, the difference between the value $24(c-a)$ of the 4d theory and the Coulomb branch dimension of the aforementioned 3d quiver (which is equal to the Higgs branch dimension of the 4d theory given by \eref{eq:HBdimDnDm} below) is equal to the total value of $24(c-a)$ of the non-Higgsable SCFTs in question. We use these two conditions as a non-trivial test of the proposed 3d mirror theories throughout the paper.

The paper is organized as follows: In Section \ref{secconv} we summarize our notation and conventions, in Section \ref{sec:confmanDnDm} we review $(D_n,D_m)$ theories and study their conformal manifold. We also compute the number of mass parameters and the dimension of the Higgs branch by counting crepant divisors. In Section \ref{sec:DpSOpsmall} we determine all the 3d mirrors of $D_p(\SO(2N))$ theories with $p<2N-2$, which is needed for the analysis of $(D_n,D_m)$ models. This result complements the analysis carried out in \cite{Carta:2021whq}. The main result is contained in Section \ref{sec:new3dmirror}, where we describe the 3d mirrors of all $(D_n,D_m)$ theories. We conclude with Appendix \ref{appendixusp} which includes new results about $D_p(\USp(2N))$ theories.

\section{Notation and convention}\label{secconv}
Throughout the paper, we use the following abbreviations in the quiver diagrams: $\SO(2N) = D_N$, $\USp(2N) = C_N$ and $\SO(2N + 1) = B_N$. We denote by $/\BZ_2$ the diagonal $\BZ_2$ quotient of the gauge symmetry.

We follow the same terminology as in \cite{Gaiotto:2008ak} to characterize the orthosymplectic gauge groups in 3d $\CN=4$ gauge theories.  A $\USp(2N)$ gauge group with $N_f$ flavors in the fundamental representation is said to be balanced, overbalanced and underbalanced if $N_f =, >, < 2N+1$, respectively. An $\SO(N)$ gauge group with $N_f$ flavors in the vector representation is said to be balanced, overbalanced and underbalanced if $N_f =, >, < N-1$, respectively.  On the contrary, in 4d $\CN=2$ gauge theories, the condition for a $\USp(2N)$ gauge group with $N_f$ to have a zero beta-function is $N_f=2N+2$, and that for an $\SO(N)$ gauge group with $N_f$ flavors to have a zero beta-function is $N_f=N-2$.  It is worth noting that a $\USp$ gauge group that satisfies the zero-beta function condition in 4d is overbalanced in 3d, whereas an $\SO$ gauge group that satisfies the zero-beta function condition in 4d is underbalanced in 3d.

We also adopt the following notations for the quiver diagrams.  
\bi
\item The $R$ copies of half-hypermultiplets in the representation $[\mathbf{2N}; \mathbf{2}]$ of the gauge group $\USp(2N)\times \SO(2)$ are denoted by
\bes{ \label{rededge}
C_{N} \begin{tikzpicture}[baseline] \draw[draw,solid,red,thick] (0,0.1)--(1,0.1) node[midway, above] {\red \scriptsize $R$}; \end{tikzpicture} D_1\fstop
}
It gives rise to an $\SU(R)$ flavor symmetry.  To make the Cartan elements of $\SU(R)$ manifest, we should interpret \eref{rededge} as denoting the half-hypermultiplets in the following representation of $\{ \USp(2N) \times \U(1) \} \times \SU(R)$, where the quantity in $\{ \cdots \}$ denotes the gauge factors and $\U(1)\cong \SO(2)$:
\bes{\label{reprededge}
[\mathbf{2N}; +1; \mathbf{\bar{R}}] \oplus [\mathbf{2N}; -1; \mathbf{R}]~. 
}

\item The $F$ flavors of hypermultiplets carrying charge $2$ under $\U(1) \cong \SO(2)$ are denoted by
\bes{ \label{wiggleline}
D_1  \begin{tikzpicture}[baseline] \draw[draw,solid,black,snake it] (0,0.1)--(1,0.1) node[midway, above] {}; \end{tikzpicture} [F]_2\coma
}
where the wiggle line and subscript 2 emphasize the charge $2$ under the $\U(1)$ gauge group.  This gives rise to an $\SU(F)$ flavor symmetry.  In other words, \eref{wiggleline} denotes the chiral multiplets in the following representation of $\U(1)  \times \SU(F)$:
\bes{
[+2; \mathbf{\bar{F}}] \oplus [-2; \mathbf{F}]~.
}
\item An edge connecting two $\SO(2)$ gauge nodes with multiplicity $M$ is denoted by  
\bes{ \label{blueedge}
D_1 \begin{tikzpicture}[baseline] \draw[draw,solid,blue,thick] (0,0.1)--(1,0.1) node[midway, above] {\blue \scriptsize $M$}; \end{tikzpicture} D_1\fstop
}
This represents $M$ copies of half-hypermultiplets in the representation $[\mathbf{2}; \mathbf{2}]$ of the gauge group $\SO(2) \times \SO(2)$. It gives rise to a $\U(M)^2/\U(1)$ flavor symmetry, whose algebra is isomorphic to $\SU(M) \times \SU(M) \times \U(1)$. To make the Cartan elements of the latter manifest, we should interpret \eref{blueedge} as denoting the half-hypermultiplets in the following representation of $\left\{\U(1) \times \U(1)\right\} \times \SU(M) \times \SU(M) \times \U(1)$, where each of the first two $\U(1)$ factors are isomorphic to each $\SO(2)$ gauge group:
\bes{ \label{repblueedge}
&[+1; +1; \bar{\mathbf{M}}; \mathbf{1}; -1] \oplus [-1; -1; \mathbf{M}; \mathbf{1}; +1 ] \\
&\oplus [+1; -1;  \mathbf{1}; \mathbf{M}; +1]\oplus [-1; +1;  \mathbf{1}; \bar{\mathbf{M}}; -1]~.  
}
\item An edge connecting an $\SO(2)$ gauge node to an $\SO(2N)$ gauge node (for $N \geq 2$) with multiplicity $M$ is denoted by  
\bes{ \label{grayedge}
D_N \begin{tikzpicture}[baseline] \draw[draw,solid,gray,thick] (0,0.1)--(1,0.1) node[midway, above] {\gray \scriptsize $M$}; \end{tikzpicture} D_1\fstop
}
This represents $M$ copies of half-hypermultiplets in the representation $[\mathbf{2N}; \mathbf{2}]$ of the gauge group $\SO(2N) \times \SO(2)$. This gives rise to an $\SU(M)$ flavor symmetry. To make the Cartan elements of $\SU(M)$ manifest, we should interpret \eref{rededge} as denoting the half-hypermultiplets in the following representation of $\{\SO(2N) \times \U(1) \} \times \SU(M)$, where the quantity in $\{ \cdots \}$ denotes the gauge factors and $\U(1)\cong \SO(2)$:
\bes{\label{reprededge}
[\mathbf{2N}; +1; \mathbf{\bar{M}}] \oplus [\mathbf{2N}; -1; \mathbf{M}]~. 
}
\ei

\section{The conformal manifold for $(D,D)$ Argyres-Douglas theories}
\label{sec:confmanDnDm}

The Calabi-Yau hypersurface singularity which engineers in Type IIB the $(D_n,D_m)$ SCFT, with $m, n\geq 3$, reads as follows
\begin{equation}
    F(u,x,y,z) = x^{n-1} + xu^2+ y^{m-1}+ yz^2 \,;\; \Omega =\frac{dudxdydz}{dF}\fstop
\end{equation} 
Because of the equivalence $(D_n,D_m)=(D_m,D_n)$ we can without loss of generality assume $m\geq n$.
By assigning scaling dimension $1$ to the holomorphic $3$-form and imposing, as usual, homogeneity of the hypersurface singularity, we can easily determine the dimension of the various coordinates: 
\be\label{dimensions}[x]=\frac{2m-2}{n+m-2}\,;\; [y]=\frac{2n-2}{n+m-2}\,;\; [u]=\frac{(n-2)(m-1)}{n+m-2}\,;\; [z]=\frac{(n-1)(m-2)}{n+m-2}\fstop\ee
The allowed deformations, which describe expectation values of Coulomb branch operators, mass parameters and relevant/marginal couplings can be parametrized as follows: 
\be\label{eqdefs} 
x^{n-1} + xu^2+ y^{m-1}+ yz^2+P(x,y)+uQ(y)+zS(x)+Muz=0\coma
\ee 
where $P$, $Q$ and $S$ are polynomials. Notice that $M$ has always dimension $1$ and is therefore a mass parameter for all values of $n$ and $m$, as can be easily seen from \eqref{dimensions}. We can also notice that the term $xy$ has always dimension $2$. 

In order to determine the dimension of the conformal manifold, we should count parameters of dimension $0$ appearing in \eqref{eqdefs}. In order to do this, it is convenient to introduce the parameters 
\be a\equiv \GCD(n-1,m-1)\,;\; p\equiv \frac{n-1}{a}\,;\; q=\frac{m-1}{a}\fstop\ee 
It is easy to see that a parameter appearing in $P(x,y)$ is marginal if and only if it multiplies a term of the form $x^{kp}y^{(a-k)q}$ with $1\leq k\leq a-1$ and therefore there are $a-1$ of them. A marginal parameter in $Q(y)$ and $S(x)$ respectively can instead appear when either $\frac{(m-1)n}{2n-2}$ or $\frac{m(n-1)}{2m-2}$ are integers. Notice that, since $n$ and $n-1$ are coprime (and analogously for $m$ and $m-1$), the two conditions above are mutually exclusive unless $n$ and $m$ are equal and even. Notice that for $m>n$ only the former can be satisfied.

In conclusion, we find that if $n=m$, the conformal manifold has dimension
\begin{equation}
    \begin{dcases}
    n-2 & \text{ if $n$ is odd,}\\
    n & \text{ if $n$ is even.}
    \end{dcases}
\end{equation}
If instead, $n< m$, we introduce
\begin{equation}
     f\equiv \frac{n(m-1)}{2(n-1)}\coma
\end{equation}
and the dimension of the conformal manifold is\footnote{The fact that for $a=1$ the theory is isolated was already noticed in \cite{Buican:2021xhs}.}
\begin{equation}
    \begin{dcases}
    a & \text{ if $f$ is an integer,}\\
    a-1 & \text{otherwise.}
    \end{dcases}
\end{equation} 

We would now like to study the cusps of the conformal manifold and identify the gauge groups becoming weakly-coupled there. In order to do that, it is convenient to set $M$ to zero in \eqref{eqdefs} and introduce, generalizing the analysis of \cite{Brandhuber:1995zp}, two Lagrange multipliers $\lambda$ and $\mu$ as follows:
\be\label{eqdefs2} 
x^{n-1} + xu^2+ y^{m-1}+ yz^2+P(x,y)+uQ(y)+zS(x)+\lambda u +\mu z=0\fstop
\ee 
Viewing this expression as a superpotential, we can integrate out the massive variables $u$ and $z$ using the equations of motion. As long as the term $Muz$ (which would couple the equations of motion) is absent, we can rewrite \eqref{eqdefs2} in terms of $x$ and $y$ only and therefore obtain a Seiberg-Witten curve describing the $(D_n,D_m)$ theory, in the limit $M\rightarrow 0$. This can be written in the form 
\be\label{SWcurve}x^{n-1}+y^{m-1}+P(x,y)+\frac{Q(y)^2}{x}+\frac{S(x)^2}{y}=0\coma\ee 
where the polynomials $P$, $Q$ and $S$ are as in \eqref{eqdefs}. 

The parametrization \eqref{SWcurve} is particularly convenient since we can plot the various deformations, which are all monomials of the form $x^ay^b$, on a plane. The coordinates of the corresponding point are given by the powers $(b,a)$. Let us give an example for ease of the reader. In the case $n=5$ and $m=7$ we can represent the allowed deformations in \eqref{SWcurve} on a plane as follows:
 \bes{\label{Polygon}
\scalebox{0.7}
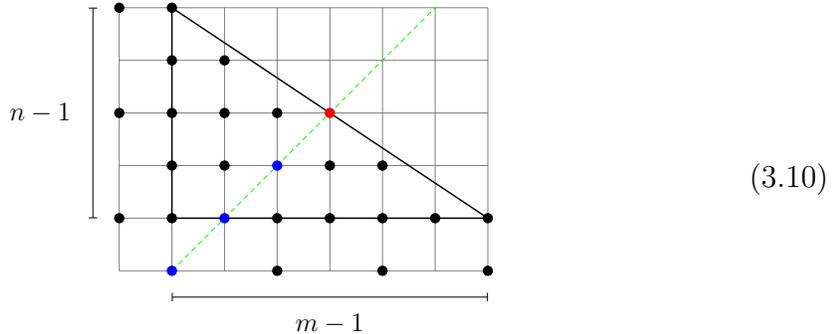
{ \begin{tikzpicture}
\def\szp{5pt};
 	\draw[gray,very thin] (0,0) grid (7,5);
 	\draw[densely dashed,green] (1,0) -- (6,5);
 	\draw[ligne,black] (1,1)--(7,1)--(1,5)--(1,1);
 	\foreach \x in {1,...,5}
 	\node[bd, minimum size=\szp] at (1,\x) {};
 	\foreach \y in {3,...,7}
	\node[bd, minimum size=\szp] at (\y,1) {};
	\node[rd, minimum size=\szp] at (4,3) {};
	\node[bld, minimum size=\szp] at (2,1) {};
	\node[bld, minimum size=\szp] at (3,2) {};
	\node[bld, minimum size=\szp] at (1,0) {};
	\node[bd, minimum size=\szp] at (0,1) {};
	\node[bd, minimum size=\szp] at (0,3) {};
	\node[bd, minimum size=\szp] at (0,5) {};
	\node[bd, minimum size=\szp] at (3,0) {}; 
	\node[bd, minimum size=\szp] at (5,0) {};
	\node[bd, minimum size=\szp] at (7,0) {};
	\node[bd, minimum size=\szp] at (2,4) {};
	\node[bd, minimum size=\szp] at (2,3) {};
	\node[bd, minimum size=\szp] at (2,2) {};
	\node[bd, minimum size=\szp] at (3,3) {};
	\node[bd, minimum size=\szp] at (4,2) {};
	\node[bd, minimum size=\szp] at (5,2) {};
	\node[] at (-1.5,3) {{\large $n-1$}};
	\draw[|-|] (-.5,1)--(-.5,5); 				
	\node[] at (4,-1) {{\large $m-1$}};
	\draw[|-|] (1,-.5)--(7,-.5); 
 \end{tikzpicture}}
 }
 
 We refer to a planar plot like \eqref{Polygon} as the Newton polygon of the singularity. All the solid dots represent allowed deformations. Those inside the triangle (including those on the boundary) appear in $P(x,y)$ in \eqref{SWcurve}. Those below the triangle arise from $Q(y)$ and those on the left come from $S(x)$. We should remember that all dots outside the triangle represent terms in \eqref{SWcurve} which are squares of more fundamental parameters. 
 
 The representation  \eqref{Polygon} is particularly convenient for identifying marginal deformations in $P(x,y)$, since they arise as dots sitting on the diagonal edge of the triangle. In the case at hand of the $(D_5,D_7)$ theory, we clearly see a single marginal parameter (denoted in red) corresponding to the deformation $x^2y^3$ in \eqref{SWcurve}. In order to identify one weakly-coupled cusp in the conformal manifold, namely providing a description of the theory involving a vector multiplet coupled to two or more matter sectors, it is convenient to recall that $[x]+[y]=2$. Therefore, if we consider a marginal deformation appearing in $P(x,y)$ of the form $x^{\alpha}y^{\beta}$ with 
 \bes{ \label{defalphabeta}
 \alpha=kp~, \quad \beta=(a-k)q~,
 }
we can then notice that the parameter multiplying the term $x^{\alpha-i}y^{\beta-i}$ has dimension $2i$, therefore suggesting these are the Casimir invariants of a gauge group, either of $\USp$ or $\SO$ type. In \eqref{Polygon} these correspond to the blue dots, which have dimension $2$, $4$ and $6$ respectively. The parameter of dimension 6, which lies outside the triangle, is actually the square of a parameter of dimension 3 and we therefore see that the dots in blue represent the Casimirs of a $\SO(6)$ group.
 
 Generalizing this observation, we are led to the conclusion that whenever we have a marginal deformation of the form $x^{\alpha}y^{\beta}$ with $\beta-\alpha$ odd, there is a $\SO(2N)$ vector multiplet (we will be more specific about the value of $N$ momentarily). If instead $\beta-\alpha$ is even, we do not find a deformation corresponding to the Pfaffian and our guess is that the gauge group is of type $\USp(2N)$. For example, in the case $n=3$ and $m=7$ we have the marginal deformation $xy^3$ and only one blue dot corresponding to a Casimir of dimension 2. There is no Casimir associated with a deformation term belonging to $Q(y)$ in \eqref{SWcurve}:
 \bes{\label{Polygon2}
\scalebox{0.7}
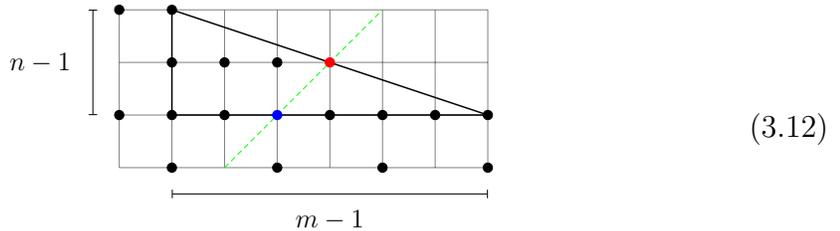
{ \begin{tikzpicture}
\def\szp{5pt};
 	\draw[step=1,gray,very thin] (0,0) grid (7,3);
 	\draw[densely dashed,green] (2,0) -- (5,3);
 	\draw[ligne,black] (1,1)--(7,1)--(1,3)--(1,1);
 	\foreach \x in {1,...,3}
 	\node[bd, minimum size=\szp] at (1,\x) {};
 	\foreach \y in {4,...,7}
	\node[bd, minimum size=\szp] at (\y,1) {};
	\node[bd, minimum size=\szp] at (2,1) {};
	\node[bd, minimum size=\szp] at (3,2) {};
	\node[bld, minimum size=\szp] at (3,1) {};
	\node[bd, minimum size=\szp] at (1,0) {};
	\node[bd, minimum size=\szp] at (0,1) {};
	\node[bd, minimum size=\szp] at (0,3) {};
	\node[bd, minimum size=\szp] at (3,0) {}; 
	\node[bd, minimum size=\szp] at (5,0) {};
	\node[bd, minimum size=\szp] at (7,0) {};
	\node[bd, minimum size=\szp] at (2,2) {};
	\node[rd, minimum size=\szp] at (4,2) {};
	\node[] at (-1.5,2) {{\large$n-1$}};
	\draw[|-|] (-.5,1)--(-.5,3); 				
	\node[] at (4,-1) {{\large $m-1$}};
	\draw[|-|] (1,-.5)--(7,-.5); 
 \end{tikzpicture}}
 }
 Regarding the matter sectors coupled to the vector multiplet, we find it convenient to exploit once again the Newton polygon and consider the straight line passing through the red and blue dots. This divides the plane in two half-planes, each containing the deformations associated with one matter sector coupled to the vector multiplet. Exploiting this guiding principle, we find that the Coulomb branch spectrum is compatible with the following description of the $(D_n,D_m)$ SCFT. 

When $\alpha-\beta$ is odd, the gauge group is of $\SO$ type, as we have already explained. More precisely, 
\begin{enumerate} 
\item For $\beta>\alpha$ we have
\begin{equation}
    D_{n-1+\beta-\alpha}(\SO(2\beta+2))- \SO(2\alpha+2) - D_{m-1+\alpha-\beta}(\SO(2\alpha+2))\fstop
    \label{eq:DnDmDecompCase1}
\end{equation}
The matter sector on the left has a partially closed regular puncture labeled by partition $\left[\beta-\alpha,\beta-\alpha,1^{2\alpha+2}\right]$. 
\item For $\beta<\alpha$ we have 
\begin{equation}
D_{n-1+\beta-\alpha}(\SO(2\beta+2))- \SO(2\beta+2) - D_{m-1+\alpha-\beta}(\SO(2\alpha+2))\fstop
\label{eq:DnDmDecompCase2}
\end{equation}
Now the matter sector on the right has a partially closed regular puncture labeled by partition $\left[\alpha-\beta,\alpha-\beta,1^{2\beta+2}\right]$.
\end{enumerate}
When instead $\alpha-\beta$ is even the gauge group is of $\USp$ type. Specifically 
\begin{enumerate}[resume]
\item For $\beta>\alpha$ we have 
\begin{equation}
    D_{n-1+\beta-\alpha}(\USp(2\beta))- \USp(2\alpha) - D_{m-1+\alpha-\beta}(\USp(2\alpha))\fstop
    \label{eq:DnDmDecompCase3}
\end{equation}
The matter sector on the left has a partially closed regular puncture labeled by partition $\left[\beta-\alpha,\beta-\alpha,1^{2\alpha}\right]$. 
\item For $\beta<\alpha$ we have 
\begin{equation}
    D_{n-1+\beta-\alpha}(\USp(2\beta))- \USp(2\beta) - D_{m-1+\alpha-\beta}(\USp(2\alpha))\fstop
    \label{eq:DnDmDecompCase4}
\end{equation}
Now the matter sector on the right has a partially closed regular puncture labeled by partition $\left[\alpha-\beta,\alpha-\beta,1^{2\beta}\right]$. 
\item For $\beta=\alpha$ we have 
\begin{equation}
    \begin{array}{ccccc}
    D_{n-1}(\USp(2\alpha))&-& \USp(2\alpha) &-& D_{m-1}(\USp(2\alpha))\fstop\\ 
                          & &        | \\ 
                          & &    [\SO(2)]
                          \end{array}
\label{eq:DnDmDecompCase5}
\end{equation}
Both matter sectors have a full regular puncture in this case, but we also have a hypermultiplet in the fundamental of $\USp(2\alpha)$.
\end{enumerate}
Notice that in all cases described above the gauging is conformal as expected. A natural question at this stage is how the parameter $M$ appearing in \eqref{eqdefs} enters the above descriptions of the $(D_n,D_m)$ theory, since it does not fit in the curve \eqref{SWcurve}. The answer is the following: In the first four cases, one of the matter sectors features a puncture of the form $[b,b,1^{c}]$. The symmetry carried by the $[1^c]$ part is always gauged, but we also have a rank $1$ factor for the global symmetry carried by the $[b,b]$ part. The corresponding mass parameter is identified with $M$. In the fifth case, instead, the $[b,b]$ part is missing and $M$ is identified with the mass of the hypermultiplet in the fundamental of the $\USp$ group. 

A careful reader might also wonder about the role of the marginal couplings appearing in $Q(y)$ or $S(x)$, since we have never discussed those. It is easy to see that $Q(y)$ includes a marginal parameter only if a term of the form $xy^{\beta}$ with $\beta$ even is marginal. According to our analysis, such a term implies the presence of a $\SO(4)$ gauge group, and therefore the marginal term in $Q(y)$ can be interpreted as providing the second marginal coupling of the $\SO(4)$ vector multiplet. Analogously, a marginal parameter in $S(x)$ implies marginality of a term of the form $x^{\alpha}y$ with $\alpha$ even, which also leads to a $\SO(4)$ gauging.

\subsection{Counting mass deformations} 

By the same token, we can compute the number of mass parameters, that is given by the deformation parameters of dimension $1$ in \eqref{eqdefs}. This equals the rank of the flavor symmetry group of $(D_n,D_m)$ theory. 

\noindent It can be checked that $Q(y)$ includes a dimension $1$ parameter only if $m-1$ is odd, and analogously $S(y)$ includes a mass parameter if $n-1$ is odd. On the other hand, $P(x,y)$ includes mass parameters if $p$ and $q$ are both odd, and in this case we find $a$ extra mass parameters. Overall, taking also into account the parameter $M$ we find the following result, depending on the parity of $n$ and $m$, we find that the number of mass parameters is: 
\begin{equation}\label{masscount} 
\begin{dcases}
1 & \text{if $n-1$ and $m-1$ are even and $p$ or $q$ is even,}\\
2 & \text{if $n-1$ is odd and $m-1$ is even,}\\
2 & \text{if $n-1$ is even and $m-1$ is odd,}\\ 
a+1 & \text{if $n-1$ and $m-1$ are even and $p$ and $q$ are odd,}\\
a+3 & \text{if $n-1$ and $m-1$ are odd.}
\end{dcases} 
\end{equation}

\subsection{Counting crepant divisors}

The dimension of the Higgs Branch of a $(G,G')$ theory can be computed adding to the number of masses, the number of crepant divisors of the canonical singularity describing the geometry of the theory \cite{Closset:2020scj, Closset:2020afy}.

The number of crepant divisors for a canonical singularity can be computed following the algorithm described in~\cite{Caibar:1999aaa,Caibar2003:aaa} that we are now going to briefly review.\footnote{We thank C. Closset for pointing out this technique to compute the number of crepant divisors.} Such algorithm can be also easily implemented, for instance, in \texttt{SageMath}~\cite{Sagemath}.\\

Let us define a polynomial 
\begin{equation}
    f= \sum_{i} a_i \prod_{j} x_j^{m_i^j}\coma
\end{equation}
where $a_i$ are integer coefficients, and $m_i=\left(m^1,\ldots, m^n\right)$ are the exponents associated to the $i$-th monomial in $f$. The locus of $f=0$ gives an isolated canonical hypersurface singularity. Let us define $M$ to be a free abelian group $\ZZ^n$, and $M_\RR=M\otimes_\ZZ \RR$. From $f$, it is possible to define its Newton polyhedron, $\Gamma_+(f)$, which is the convex hull in $M_\RR$ of the set given by the union of the $m_i$ and the positive quadrant in $\RR$. \\
We now introduce a set of vectors $\alpha=(\alpha_1,\ldots \alpha_n)$ that belongs to $N=\text{Hom}_\ZZ(M,\ZZ)$, which is the dual of $M$, and for each $\alpha$, we define $\alpha(m)=\alpha\cdot m_i$ and
\begin{equation}
    \alpha(f)=\min_{m_i\in f} \alpha \cdot m_i\fstop
\end{equation}
The vectors $\alpha$ are called \textit{weightings}, and from this set of vectors, we call \textit{crepant weightings} those that satisfies
\begin{equation}
    |\alpha| = \alpha(f)+1\coma
\end{equation}
where $|\alpha|$ is the sum of the components of $\alpha$.

For each crepant weighting we also introduce 
\begin{equation}
    g(\alpha)=\min_{m_i\in \Gamma_+(f)} \alpha\cdot m_i\coma
\end{equation}
and we call the face of $\Gamma(f)$ corresponding to $\alpha$ as the set
\begin{equation}
    \Gamma_\alpha=\left\{m_i\in \Gamma_+(f)\,:\,\alpha(m)=g(\alpha)\right\}\fstop
\end{equation}
Finally, we define the length $\Gamma_\alpha$ as the number of vectors composing the face $\Gamma_\alpha$ minus $1$. The number of crepant divisors for an isolated canonical singularity $X$ is~\cite{Caibar:1999aaa,Caibar2003:aaa}
\begin{equation}
    c(X)=\sum_\alpha c(\alpha) \text{ with } c(\alpha)=\begin{dcases}
    \text{length } \Gamma_\alpha & \text{if }\dim \Gamma_\alpha =1 \coma\\
    1 & \text{if }\dim \Gamma_\alpha \geq 2\fstop
    \end{dcases}
\end{equation}
The HB of a $(D_n,D_m)$ AD theory is given by the sum of the number of masses in~\eqref{masscount} and the number of crepant divisors. We find that for given a $(D_n,D_m)$ theory $X$, for $m\geq n$, the number of its crepant divisors is
\begin{equation}
    c(X)=\left\lfloor \frac{n}{2}\right\rfloor-1\fstop
\end{equation}
Therefore, the HB dimension for a $(D_n,D_m)$, using Eq.~\ref{masscount}, can be written as follows:
\begin{equation}\label{eq:HBdimDnDm} 
\begin{dcases}
\left\lfloor \frac{n}{2}\right\rfloor & \text{if $n-1$ and $m-1$ are even and $p$ or $q$ is even,}\\
\left\lfloor \frac{n}{2}\right\rfloor+1 & \text{if $n-1$ is odd and $m-1$ is even,}\\
\left\lfloor \frac{n}{2}\right\rfloor+1 & \text{if $n-1$ is even and $m-1$ is odd,}\\
a+\left\lfloor \frac{n}{2}\right\rfloor & \text{if $n-1$ and $m-1$ are even and $p$ and $q$ are odd,}\\
a+\left\lfloor \frac{n}{2}\right\rfloor+2 & \text{if $n-1$ and $m-1$ are odd.}
\end{dcases} 
\end{equation}

\section{The 3d mirror of $D_p(\SO(2N))$ theories with $p<2N-2$}\label{sec:DpSOpsmall} 

In \cite{Carta:2021whq} we have determined the 3d mirror of $D_p(\SO(2N))$ theories with $p < 2N-2$ only in some cases. Here we would like to fill in this gap since this will be useful for determining the 3d mirrors of $(D_n,D_m)$ theories. We will divide the analysis in two cases; $p$ odd and $p$ even. The analysis will be largely based on an analogy with the $\SU(N)$ case, which is well understood and is discussed in detail in \cite{Giacomelli:2020ryy}.

\subsection{$D_p(\SO(2N))$ theories with $p$ odd} 
\label{sec:DpSOpodd}

These models are the simplest since they do not include any mass parameters apart from those associated with the $\SO(2N)$ global symmetry. It turns out that in this case the 3d mirror is given by two orthosymplectic quiver tails with flavors attached at some of the gauge nodes, which are all balanced. 

For convenience, we define the parameter $x$ as follows:
\bes{
x\equiv\left\lfloor\frac{2N-2}{p}\right\rfloor~.
}

For $x$ odd or for $2N-2$ divisible by $p$ (where, for the latter, $x$ is even),
we propose that the 3d mirror of $D_p(\SO(2N))$ is identified with the theory $T_{\left[1^{2N}\right]}^{\sigma}[\SO(2N)]$, where 
\be \label{defsigma}
\sigma=\left[(x+1)^{2N-1-px},x^{xp+p+1-2N},1\right]\fstop
\ee
Notice that the above $\sigma$ is always an even partition of $2N$ regardless of the parity of $x$. If $2N-2$ is divisible by $p$, then $\sigma$ can be written as $\sigma =[x+1, x^{p-1}, 1]$.  The explicit quivers can be described as follows: 
\be\label{mirrorXodd}\begin{array}{ccccccccccccccccc} 
& &  & & &  &  &  & \left[C_{\frac{2N-1-px}{2}}\right] & & & & & & & & \\
& &  & & &  &  &  & | & & & & & & & & \\
C_{\frac{p-1}{2}} & - & B_{p-1} & - & \cdots & - & C_{\frac{p-1}{2}x} & - & D_{\frac{2N-1-x}{2}} & - & C_{\frac{2N-3-x}{2}} & - & \cdots & - & C_1 & - & D_1 \\
| & &  & & &  & | &  & & & & & & & & & \\ 
 B_0 & & & & & & \left[B_{\frac{xp+p-2N}{2}}\right] & & & & & & & & & & \\ 
\end{array}\ee 
The tail on the left has alternating gauge groups of type $C$ and $B$, with increasing ranks, which are multiples of $\frac{p-1}{2}$. The tail on the right is instead equivalent to $T[\SO(2N-1-x)]$.  The mirror theory also comes with
\begin{equation} \label{Hfreepodd}
    H_{\text{free}}=X\left(\frac{p-1}{2}-X\right) \coma
\end{equation}
free hypermultiplets,\footnote{We remark that the expression for the number of the free hypermultiplets $H_{\text{free}}$ in the mirror theory for $D_p(\SO(2N))$ with $p$ odd and $p\geq 2N-2$, given by \cite[(5.4)]{Carta:2021whq}, can also be written as $H_{\text{free}}= \left(\frac{p-2N+1}{2} \right) N = \left[\frac{p-1}{2}-(N-1)\right]N$. We see that \eref{Hfreepodd} indeed has a similar form to such an expression. \label{foot1}} where
\begin{equation}
    X =
    \left( N+\frac{x-1}{2} \right) \bmod \left(\frac{p+1}{2} \right)~.
\end{equation}

Let us discuss the case in which $p$ divides $2N-2$. Here $X =0$ and there are no free hypermultiplets. Since $p$ is odd, $x = (2N-2)/p$ is even and, according to \cite[Appendix C]{Cecotti:2013lda}, the corresponding $D_p(\SO(2N))$ theory in $4$d admits a Lagrangian description. As we commented in \cite[Section 8.4]{Carta:2021whq}, the Lagrangian description of the 4d theory is the same as that of the quiver description of the 3d $T_{\sigma}[\SO(2N)]$ theory, with $\sigma$ given by \eref{defsigma}, namely 
\bes{
[D_{\fm p+1}]-C_{\fm(p-1)}-D_{\fm(p-2)+1}-\cdots-C_{2\fm}-D_{\fm+1}
}
where $\fm \equiv x/2$. Each $C$ and $D$ gauge group in the 4d quiver theory has zero beta-function, whereas in the 3d quiver the $C$-gauge group is overbalanced and the $D$ gauge group is underbalanced, rendering the theory ``bad'' in the sense of \cite{Gaiotto:2008ak}. At this stage, it is not clear whether the quiver for $T_{\sigma}[\SO(2N)]$ describes the reduction of such a $D_p(\SO(2N))$ theory to 3d.  If we {\it assume}\footnote{This assumption should be taken cautiously. We observe that only when $p$ divides $2N-2$, the Coulomb branch dimension of the mirror theory is $(p-1)/2$ larger than the value $24(c-a)$ of the corresponding 4d theory. We have checked that all the other cases do not have this problem, namely the Coulomb branch dimension of the mirror theory is always less than or equal to the integer part of $24(c-a)$.  In this case, it is not clear whether $24(c-a)$ is equal to the Higgs branch of the 4d theory, since the orthogonal gauge groups may not be completely Higgsed at a generic point on the Higgs branch. This is due to the fact that the orthogonal gauge group which is conformal in 4d is underbalanced in 3d.  Nevertheless, we emphasize that the Higgs branch dimension of the mirror theory \eref{defsigma} is exactly equal to the rank of the corresponding 4d theory, as it should be.\label{caution} } that this is true, then the mirror theory is as described above. We hope to gain better understanding of this case in the future.

For $x$ even and $2N-2$ not divisible by $p$, we propose that the 3d mirror for $D_p(\SO(2N))$ is identified with the theory $T^{\sigma'}_{[1^{2N}]}[\SO(2N)]$, where the partition $\sigma'$ can be obtained from the partition $\sigma$ by ``lifting a box up'', \ie~
\bes{
\sigma' = [(x + 1)^{2 N - p x} , x^{x p + p - 1 - 2 N} , x - 1, 1]~.
}
Note that if the box is not liftable, i.e. $x p + p - 1 - 2 N < 0$, the partition $\sigma'$ is identical to the partition $\sigma$, e.g. $N=5,\, p=3$, we have $\sigma = \sigma' = [3, 3, 3, 1]$.  The theory $T^{\sigma'}_{[1^{2N}]}(\SO(2N))$ admits a quiver description of the form 
\be\label{mirrorXeven}
\scalebox{0.9}{$
\begin{array}{ccccccccccccccccccc} 
& &  & & & & &  & \left[C_{\frac{px+p-1-2N}{2}}\right]  & & & & & & & & & & \\
& &  & & & & & & |  & & & & && & & & \\
C_{\frac{p-1}{2}} & - & B_{p-1}   & - & \cdots  &- & C_{\frac{p-1}{2}(x-1)} &- &D_{\frac{p -1}{2} x}& - & C_{\frac{2N-2-x}{2}} & - & D_{\frac{2N-2-x}{2}} & - & \cdots & - & C_1 & - & D_1 \\
| & &  & & &  &| & & & & | & & & & & & & & \\ 
 B_0 & & & & & & B_0 & & & & \left[D_{\frac{2N-px}{2}}\right] & & & & & & & & \\ 
\end{array}$}
\ee 
The number of free hypermultiplets in this case (\ie~ $x$ even and $p$ does not divide $2N-2$) is
\bes{
H_{\text{free}} = Y \left( \frac{p+1}{2} - Y \right)\coma
}
where 
\bes{
Y = \left( N+ \frac{x}{2} \right) \bmod \left(\frac{p+1}{2}\right)~.
}

Let us mention some consistency checks for the proposed mirror theory.  It is convenient to discuss these via examples.
\bi
\item {\bf The case of $N=9$ and $p=7$}.  Here $x=2$.  The 4d theory has $24(c-a) = 459/7 = 65+(4/7)$ and it has rank $27$.  The proposed mirror theory has CB dimension $65$ and HB dimension $23$.  We thus have $27-23=4$ free hypermultiplets, in agreement with the fact that the non-Higgsable SCFT has $24(c-a)$ equal to $4/7$, which is expected to be that of the $(A_1, A_4)^{\otimes 2}$ theory whose rank is $(2 \times 2)=4$.
\item {\bf The case of $N=12$ and $p=5$}.  Here $x=4$.  The 4d theory has $24(c-a) = 552/5 = 110+(2/5)$ and it has rank $24$.  The proposed mirror theory $T^{\sigma'}[\SO(2N)]$ has CB dimension 110 and HB dimension 22.  We thus have $24-22=2$ free hypermultiplets, in agreement with the fact that the non-Higgsable SCFT has $24(c-a)$ equal to $2/5$, which is expected to be that of the $(A_1, A_2)^{\otimes 2}$ theory whose rank is $(2 \times 1)=2$.
\item {\bf The case of $N=12$ and $p=7$}.  Here $x=3$.  The 4d theory has $24(c-a) = 828/7 = 118+(2/7)$ and it has rank $36$.  The proposed mirror theory $T^{\sigma}(\SO(2N))$ has CB dimension $118$ and HB dimension $34$.  We thus have $36-34=2$ free hypermultiplets, in agreement with the fact that the non-Higgsable SCFT has $24(c-a)$ equal to $2/7$, which is expected to be that of the $(A_1, A_4)$ theory whose rank is $2$.
\ei

Let us also briefly comment on the procedure of ``lifting a box up'' in the case of $x$ even and $2N-2$ not divisible by $p$.  The reason for this is two-fold. First, this procedure leads to the desired Higgs branch and Coulomb branch dimensions of the mirror theory that pass the above checks.  Secondly, this procedure cures the ``badness'' of the $T_\sigma[\SO(2N)]$ theory (see also \cite{Gaiotto:2012uq} for a related discussion). To illustrate this point, we take an example of $N=12$ and $p=5$, where $\sigma=[5,5,5,4,4,1]$, $\sigma'=[5,5,5,5,3,1]$, and
\bes{
T_\sigma[\SO(24)]: &\quad [D_{12}]-C_9-D_7-C_4-D_3 \\
T_{\sigma'} [\SO(24)]: &\quad [D_{12}]-C_9-D_7-C_4-D_2 \\
}
Observe that the former contains an underbalanced $D_3$ gauge node, rendering the theory bad, whereas the latter does not contain any underbalanced gauge nodes.

\subsection{$D_p(\SO(2N))$ theories with $p$ even} 

The difference with respect to the previous case is that for $p$ even, we have extra mass parameters besides those associated with the $\SO(2N)$ global symmetry. Their number depends on the parity of $\frac{2N-2}{\GCD(2N-2,p)}$ and we will therefore consider the two cases separately.

\subsubsection{The case $\frac{2N-2}{\GCD(2N-2,p)}$ even}
\label{sec:DpSOpeven-ratioeven}

In this case we have only one mass parameter besides those of $\SO(2N)$. The 3d mirror is given by a orthosymplectic quiver with $2N-2$ balanced gauge groups and one overbalanced $D_1$ group whose FI parameter accounts for the extra mass parameter of the 4d theory. We introduce again the parameter 
\bes{
x=\left\lfloor\frac{2N-2}{p}\right\rfloor
}
and we discuss the cases in which $p$ does and does not divide $2N-2$ separately. 

If $p$ does not divide $2N-2$ (here $x$ can be even or odd), we propose that the 3d mirror is given by a $T_{\rho}^{\sigma}[\SO(2N+2p-2)]$ theory with 
\be \label{partmain}
\rho=[(p-1)^2,1^{2N}]\,;\; \sigma=[(x+3)^{2N-2-xp},(x+2)^{xp+p-2N+2}]\fstop\ee 
For $x$ odd, the quiver reads 
\be\label{mirrorXodd2}\begin{array}{ccccccccccccccccc} 
& &  & & & & \left[D_{\frac{xp+p-2N+2}{2}}\right] & & & &  & & & & & & \\
& &  & & & & | & & & & & & & & & & \\
D_1 & - & C_{p/2} & - & \cdots & - & C_{xp/2-(x-1)/2} & - & D_{N-1-(x-1)/2} & - & C_{\frac{2N-3-x}{2}} & - & \cdots & - & C_1 & - & D_1 \\
 & &  & & &  & & & | & & & & & & & & \\ 
 & & & & & & & & \left[C_{\frac{2N-2-xp}{2}}\right] & & & & & & & & \\ 
\end{array}\ee 
Again, all the nodes not indicated explicitly are of C or D type and balanced. On the other hand, for $x$ even, the quiver reads
\bes{ \label{xevenaaa}
\begin{array}{ccccccccccccccccc} 
& &  & & & & & & & & \left[D_{\frac{2N-2-xp}{2}}\right] & & & & & & \\
& &  & & & && & & & | & & & & & & \\
D_1 & - & C_{p/2} & - & \cdots & - & C_{(x-1)p/2-(x-2)/2} & - & D_{x(p-1)/2+1} & - & C_{\frac{2N-2-x}{2}} & - & \cdots & - & C_1 & - & D_1 \\
 & &  & & &  & & & | & & & & & & & & \\ 
 & & & & & & & & \left[C_{\frac{px+p-(2N-2)}{2}}\right] & & & & & & & & \\ 
\end{array}
}
This mirror theory also comes with a number of free hypermultiplets given by\footnote{We remark that the expression for the number of the free hypermultiplets $H_{\text{free}}$ in the mirror theory for $D_p(\SO(2N))$ with $p$ even, $p\geq 2N-2$ and $\frac{2N-2}{\GCD(2N-2,p)}$ even, given by \cite[(6.4)]{Carta:2021whq}, can also be written as $H_{\text{free}}= \left(\frac{p}{2}-N+\frac{5}{2} \right) N -p -\frac{3}{2}$. We see that \eref{HfreeDpSO2Npevenratioevenxodd}, for $x$ even, indeed has the same form as the former expression with $N$ replaced by $N_p$.}
\bes{ \label{HfreeDpSO2Npevenratioevenxodd}
H_{\text{free}} = \begin{cases}
(p-N_p+2) \left(N_p-\frac{p}{2}+\frac{1}{2}\right)-p-\frac{3}{2}~, & \qquad \text{if $x$ odd,} \\
 \left( \frac{p}{2} - N_p + \frac{5}{2} \right)  N_p - p - \frac{3}{2}~, & \qquad \text{if $x$ even,}
\end{cases}
}
with $N_p$ defined as 
\bes{\label{defnp}
N_p = N \,\, \mod \,\, p~.
}

If $p$ divides $2N-2$,\footnote{In this case $x=\frac{2N-2}{p}$ is even by our assumption that $\frac{2N-2}{\GCD(2N-2,p)} = \frac{2N-2}{p}=x$ is even.} the 4d theory $D_p(\SO(2N))$ admits a Lagrangian description, which turns out to coincide with the quiver for $T^{\rho}_{\sigma}[\SO(2N+2p-2)]$ theory, with
\bes{ \label{partA}
\rho&=[(p-1)^2,1^{2N}] = [(p-1)^2,1^{xp+2}] ~;\\ \sigma&=[(x+3)^{2N-1-xp},(x+2)^{xp+p-2N},x+1]
= [x+3,(x+2)^{p-2},x+1]
}
namely
\bes{
[D_{\fm p+1}]-C_{\fm(p-1)}-D_{\fm(p-2)+1}-\cdots-D_{2\fm+1}-C_\fm-[D_1]
}
where $\fm \equiv x/2$. The 3d mirror of this quiver is given by a $T_{\rho}^{\sigma}[\SO(2N+2p-2)]$ theory
\be\label{mirrorXeven2}\begin{array}{ccccccccccccccccc} 
& &  & & & & B_0 & & & & B_0 & & & & & & \\
& &  & & & & | & & & & | & & & & & & \\
D_1 & - & C_{p/2} & - & \cdots & - & C_{\frac{(x-1)p}{2}-\frac{x-2}{2}} & - & B_{\frac{x(p-1)}{2}} & - & C_{\frac{x(p-1)}{2}} & - & \cdots & - & C_1 & - & D_1 \\
 & &  & & &  & & & | & & & & & & & & \\ 
 & & & & & & & & \left[C_{\frac{p-2}{2}}\right] & & & & & & & & \\ 
\end{array}\ee 
All the gauge groups not indicated explicitly are alternating of $C$ and $D$ type. The ranks are fixed by requiring all groups to be balanced. It can be checked that in this case, the Higgs branch dimension of the mirror theory \eref{mirrorXeven2} is exactly equal to the rank of the corresponding 4d theory $D_p(\SO(2N))$.  We conclude that there are no free (twisted) hypermultiplets upon reduction of such a 4d theory on a circle.  Similarly to Footnote \ref{foot1}, we remark that the Coulomb branch dimension of the mirror theory \eref{mirrorXeven2} is $(p-2)/2$ larger than the value of $24(c-a)$ of the corresponding 4d theory.  In the special case of $p=2$, the 4d theory is simply the $\USp(x)$ SQCD with $x+2$ flavors.  The corresponding 3d mirror theory \eref{mirrorXeven2}, with $p=2$, turns out to be coincident with \cite[Figure 11]{Feng:2000eq} with $k=x/2$ and $N=x+2$, as it should be. 

We can perform a similar consistency check for the proposed mirror theory, as in the previous subsection.
\bi
\item {\bf The case of $N=7$ and $p=10$}.  Here $x=1$.  The 4d theory has $24(c-a) = 211/5 = 42+(1/5)$ and it has rank $31$.  The proposed mirror theory $T_{\rho}^{\sigma}[\SO(2N+2p-2)]$ has CB dimension $42$ and HB dimension $30$.  We thus have $31-30=1$ free hypermultiplets, in agreement with the fact that the non-Higgsable SCFT has $24(c-a)$ equal to $1/5$, which is expected to be that of the $(A_1, A_2)$ theory whose rank is $1$.
\item {\bf The case of $N=13$ and $p=10$}.  Here $x=2$. The 4d theory has $24(c-a) = 742/5 = 148+(2/5)$ and it has rank $58$.  The proposed mirror theory $T_{\rho}^{\sigma}[\SO(2N+2p-2)]$ has CB dimension $148$ and HB dimension $56$.  We thus have $58-56=2$ free hypermultiplets, in agreement with the fact that the non-Higgsable SCFT has $24(c-a)$ equal to $2/5$, which is expected to be that of the $(A_1, A_2)^{\otimes 2}$ theory whose rank is $2$.
\ei

\subsubsection{The case $\frac{2N-2}{\GCD(2N-2,p)}$ odd}
\label{sec:DpSOpeven-ratioodd}

We recall that the only way in which we used the fact that $\frac{2N-2}{\GCD(2N-2,p)}$ is even in Section \ref{sec:DpSOpeven-ratioeven} is in requiring that the flavors attached to the central $C$-type node in \eref{mirrorXodd2} and \eref{xevenaaa} are not gauged. This has to be the case in order to match the rank of the global symmetry of the 4d theory discussed in Section \ref{sec:DpSOpeven-ratioeven}. 

We shall shortly see that the main part of the mirror quiver in the previous section can be carried through in the case of $\frac{2N-2}{\GCD(2N-2,p)}$ odd as well. We claim that the structure of the two tails in the mirror dual is the same as in Section \ref{sec:DpSOpeven-ratioeven}, which is also supported by the known form of the mirrors presented in \cite[Section 8.3]{Carta:2021whq}. The only differences are that 
\ben
\item The $D$-type flavor nodes in \eref{mirrorXodd2} and \eref{xevenaaa} are now gauged to account for the larger number of mass parameters and they form a complete graph, whose structure we can try to guess using as a guidance the special cases we have already worked out.
\item The $C$-type flavor node that is attached to a central $D$-type gauge node in \eref{mirrorXodd2} and \eref{xevenaaa} is replaced by lines with appropriate multiplicities that connect such a $D$-type gauge node to the vertices of the complete graph.
\een
Let us now be explicit about the above description.

As before, we use the notation
\bes{
x=\left\lfloor\frac{2N-2}{p}\right\rfloor~.
}
We start from $T_{\rho}^{\sigma}[\SO(2N+2p-2)]$ theory with the same partitions as in \eref{partmain}, namely
\be \label{mainpartitionsa} \rho=[(p-1)^2,1^{2N}]\,;\; \sigma=[(x+3)^{2N-2-xp},(x+2)^{xp+p-2N+2}]\fstop\ee 
Let us first consider the case $x$ odd.  The quiver description of such a theory is given by
\eqref{mirrorXodd2}.  For convenience, we reproduce the diagram here:
\be\label{mirrorXodd3}\begin{array}{ccccccccccccccccc} 
& &  & & & & \left[D_{\frac{xp+p-2N+2}{2}}\right] & & & &  & & & & & & \\
& &  & & & & | & & & & & & & & & & \\
D_1 & - & C_{p/2} & - & \cdots & - & C_{xp/2-(x-1)/2} & - & D_{N-1-(x-1)/2} & - & C_{\frac{2N-3-x}{2}} & - & \cdots & - & C_1 & - & D_1 \\
 & &  & & &  & & & | & & & & & & & & \\ 
 & & & & & & & & \left[C_{\frac{2N-2-xp}{2}}\right] & & & & & & & & \\ 
\end{array}\ee 
Since we should now introduce $\GCD(2N-2,p)/2$ new mass parameters, we replace the $\left[D_{\frac{xp+p-2N+2}{2}}\right]$ flavor node with a collection of $\GCD(2N-2,p)/2$ $D_1$ gauge nodes. We assume they are all identical. By analogy with \cite[Section 5]{Giacomelli:2020ryy}, we propose the following candidate structure for the mirror theory:
\begin{enumerate} 
\item Keep the middle line of quiver \eref{mirrorXodd3} as it is.
\item Replace the $\left[D_{\frac{xp+p-2N+2}{2}}\right]$ flavor node with a collection of $\GCD(2N-2,p)/2$ $D_1$ gauge nodes. \label{step1}
\item Each $D_1$ group is connected to the node $C_{xp/2-(x-1)/2}$ with an edge of multiplicity 
\bes{
m_A \equiv \frac{(x+1)p-(2N-2)}{\GCD(2N-2,p)}~.
}
\item Remove the $[C_{\frac{2N-2-xp}{2}}]$ from \eref{mirrorXodd3} and connect the $D_{N-1-(x-1)/2}$ gauge node to each of the $D_1$ gauge nodes in Step \ref{step1} by an edge with multiplicity
\bes{
m_B \equiv \frac{(2N-2)-xp}{\GCD(2N-2,p)}~.
}
\item Each pair of $D_1$ nodes is connected by an edge whose multiplicity is equal to 
\bes{ \label{deffrakk}
m_G \equiv m_A m_B= \frac{[(x+1)p-(2N-2)][(2N-2)-xp]}{\GCD(2N-2,p)^2}~.
}
These form a complete graph of $\GCD(2N-2,p)/2$ nodes, with all edge multiplicity equal to $m_G$.
\item There are 
\bes{
\frac{1}{2}(m_A-1)m_B
} 
hypermultiplets of charge 2 charged under each $\U(1)\cong D_1$ gauge group.
\item The quiver constructed above has an overall $\BZ_2$ that needs to be decoupled.
\end{enumerate}

Assuming there are no other ingredients, the dimension of the HB of the quiver matches the dimension of the CB of the 4d SCFT provided that the number of free hypermultiplets is 
\bes{ \label{HfreeDpSO2N}
\scalebox{0.9}{$
\begin{split}
H_{\text{free}}
&= \frac{1}{4} (m_A-1)(m_B-1) \GCD(2N-2,p) \\
&= \frac{\left[(x+1) p -(2N-2)-\GCD(2N-2,p) \right] \left[(2N-2)- xp -\GCD(2N-2,p) \right]}{4 \GCD(2N-2,p)}~.
\end{split}
$}
}
It is worth pointing out that the above expression of $H_{\text{free}}$ takes the same form as that for the $D_p(\SU(N))$ theory with $p \leq N$, where the latter is given by \cite[(5.5)]{Giacomelli:2020ryy}. The only difference between the two expressions are the prefactors, where they are $1/4$ in the former and $1/2$ in the latter.

We now turn to the case $x$ even. The procedure is very similar to that of the case $x$ odd, with the roles of $m_A$ and $m_B$ interchanged. Explicitly, we start from the theory \eref{mainpartitionsa}, whose quiver description is given by \eref{xevenaaa}. For convenience, we reproduce it here again:
\be\label{mirrorXodd4}
\begin{array}{ccccccccccccccccc} 
& &  & & & & & & & & \left[D_{\frac{2N-2-xp}{2}}\right] & & & & & & \\
& &  & & & && & & & | & & & & & & \\
D_1 & - & C_{p/2} & - & \cdots & - & C_{(x-1)p/2-(x-2)/2} & - & D_{x(p-1)/2+1} & - & C_{\frac{2N-2-x}{2}} & - & \cdots & - & C_1 & - & D_1 \\
 & &  & & &  & & & | & & & & & & & & \\ 
 & & & & & & & & \left[C_{\frac{px+p-(2N-2)}{2}}\right] & & & & & & & & \\ 
\end{array}\ee 
We then follow the subsequent steps:
\ben
\item Keep the middle line of quiver \eref{mirrorXodd4} as it is.
\item Replace the $\left[D_{\frac{2N-2-xp}{2}}\right]$ flavor node with a collection of $\GCD(2N-2,p)/2$ $D_1$ gauge nodes. \label{step1a}
\item Each $D_1$ group is connected to the node $C_{(2N-2-x)/2}$ with an edge of multiplicity $m_B$.
\item Remove the $\left[C_{\frac{p x+p-(2N-2)}{2}}\right]$ from \eref{mirrorXodd4} and connect the $D_{x(p-1)/2+1}$ gauge node to each of the $D_1$ gauge nodes in step \ref{step1a} by an edge with multiplicity $m_A$.
\item Each pair of $D_1$ nodes is connected by an edge whose multiplicity is equal to $m_G$.
These form a complete graph of $\GCD(2N-2,p)/2$ nodes, with all edge multiplicity equal to $m_G=m_A m_B$.
\item There are 
\bes{
\frac{1}{2}(m_B-1)m_A
} 
hypermultiplets of charge 2 charged under each $\U(1)\cong D_1$ gauge group.
\item The quiver constructed above has an overall $\BZ_2$ that needs to be decoupled.
\een
There are also $H_{\text{free}}$ free hypermultiplets, given by \eref{HfreeDpSO2N}.

\subsubsection*{Examples}
\bi
\item Let us consider the case in which $p$ divides $2N-2$, so that
\bes{
2N-2 = x p~, \quad \text{$x$ is odd}~.
}
The mirror theory for $D_p(\SO(x p+2))$ is then
\bes{
\begin{tikzpicture}[baseline,font=\small]
\node[draw=none] (C1) at (-3,0) {$C_{\frac{xp-x+1}{2}}$};
\node[draw=none, right of =C1,xshift=1cm] (D2N){$D_{\frac{xp-x+1}{2}}$};
\node[draw=none, left of =C1,xshift=-1.5cm] (DL){$D_1-C_{p/2}-\cdots$};
\node[draw=none, right of = D2N, xshift=3cm] (D10) {$C_{\frac{px-x-1}{2}}-D_{\frac{px-x-1}{2}}-\cdots-C_1-D_1$};
\node[draw=none, above left of=C1, yshift=0.5cm] (D1a) {$D_1$};
\node[draw=none, above right of=C1,  yshift=0.5cm] (D1b) {$D_1$};
\node[draw=none, above of=C1, yshift=0.2cm] (dots) {\large $\cdots$};
\draw[solid] (DL)--(C1)--(D2N)--(D10);
\draw[solid] (C1) -- (D1a) ;
\draw[solid] (C1) -- (D1b);
\draw [decorate,decoration={brace,amplitude=10pt,raise=4pt},yshift=-0.8cm]
(-4,2) -- (-2,2) node [black,midway,above,yshift=0.4cm] {\footnotesize
$p/2$ nodes};
\end{tikzpicture} \quad /\BZ_2
}
When $x=1$, namely $p=2N-2$, we recover the quiver described in \cite[Section 6.2]{Carta:2021whq} as expected.
\item The mirror theory of $D_{4\fN}(\SO(4\fN+4))$, for which we have $N=2\fN+2$, $2N-2=4\fN+2$, $p=4\fN$, $\GCD(2N-2,p)=2$, $x=1$, $m_A=2\fN-1$, $m_B=1$, is described by\footnote{This quiver has
\bes{ \nn
\dim_\BH \text{HB}\, \eref{mirr4NSO4Np4a} = 4 \fN^2+3 \fN-2~, \\
\dim_\BH \text{CB}\, \eref{mirr4NSO4Np4a} = 4 \fN^2+6 \fN+3~,
}
which are equal to the CB and HB dimensions of the 4d $D_{4\fN}(\SO(4\fN+4))$ theory, respectively.  Moreover, it has the CB symmetry $\SO(4\fN+4) \times \SO(2)^2$, which is the same as the flavor symmetry of the 4d theory.}
\bes{ \label{mirr4NSO4Np4a}
\begin{tikzpicture}[baseline]
\node[draw=none] (D10) at (6,0) {$C_{2\fN}-D_{2\fN}-\cdots-C_1-D_1$};
\node[draw=none] (D2N) at (2.2,0) {$D_{2 \fN+1}$};
\node[draw=none] (C1) at (0.8,0) {$C_{2\fN}$};
\node[draw=none] (F) at (-1,1) {$[\fN-1]_2$};
\node[draw=none] (D1a) at (0.8,1) {$D_1$};
\node[draw=none] (D1b) at (-1,0){$D_1$};
\draw[solid, thick, snake it] (F) to (D1a);
\draw[solid] (C1)--(D2N)--(D10);
\draw[solid,red,thick] (C1) -- node[midway,left] {\scriptsize $2\fN-1$} (D1a) ;
\draw[solid] (C1) to (D1b);
\draw[solid,gray, thick] (D2N) -- node[midway,above right] {\scriptsize $1$} (D1a);
\end{tikzpicture} \qquad /\BZ_2
}
There are no free hypermultiplets in this example.
\item The mirror theory of $D_{24}(\SO(90))$, for which we have $N=45$, $2N-2=88$, $p=24$, $\GCD(2N-2,p)=8$, $x=3$, $m_A=1$, $m_B=2$, is described by
\bes{
\scalebox{0.85}{
\begin{tikzpicture}[baseline]
\node[draw=none] (C1) at (0,0) {$C_{35}$};
\node[draw=none, right = of C1] (D10)  {$D_{43}$};
\node[draw=none, right = of D10] (D11) {$C_{42}-D_{42}-\cdots-C_1-D_1$};
\node[draw=none] (D1a) at (-1,2) {$D_1$};
\node[draw=none] (D1b) at (1,2) {$D_1$};
\node[draw=none] (D1c) at (-1,3) {$D_1$};
\node[draw=none] (D1d) at (1,3) {$D_1$};
\node[draw=none, left = of C1] (D1L1) {$D_{24}$};
\node[draw=none, left = of D1L1] (D1L2) {$C_{12}$};
\node[draw=none, left = of D1L2] (D1L3) {$D_1$};
\draw[solid] (C1)--(D10)--(D11);
\draw[solid,red,thick] (C1) -- node[midway,below left] {\scriptsize $1$} (D1a) ;
\draw[solid,red,thick] (C1) -- (D1b) ;
\draw[solid,red,thick] (C1) -- (D1c) ;
\draw[solid,red,thick] (C1) -- (D1d) ;
\draw[solid,blue,thick] (D1a)--(D1b) ;
\draw[solid,blue,thick] (D1a)--(D1c) ;
\draw[solid,blue,thick] (D1a)--(D1d) ;
\draw[solid,blue,thick] (D1b)--(D1c) ;
\draw[solid,blue,thick] (D1b)-- (D1d) ;
\draw[solid,blue,thick] (D1c)-- node[midway,above] {\scriptsize $2$} (D1d) ;
\draw[solid] (D1L3)--(D1L2)--(D1L1)--(C1);
\draw[solid, thick, gray] (D10) -- (D1a);
\draw[solid, thick, gray] (D10) -- (D1b);
\draw[solid, thick, gray] (D10) -- (D1c);
\draw[solid, thick, gray] (D10) -- node[midway,right] {\scriptsize $2$}  (D1d);
\end{tikzpicture} 
} \quad /\BZ_2 }
There are no free hypermultiplets in this example.
\item The mirror theory of $D_{30}(\SO(80))$, for which we have $N=40$, $2N-2=78$, $p=30$, $\GCD(2N-2,p)=6$, $x=2$, $m_A=2$, $m_B=3$, is described by
\bes{
\scalebox{0.85}{
\begin{tikzpicture}[baseline]
\node[draw=none] (C1) at (0,0) {$C_{38}$};
\node[draw=none, right = of C1] (D10)  {$D_{38}$};
\node[draw=none, right = of D10] (D11) {$C_{37}-D_{37}-\cdots-C_1-D_1$};
\node[draw=none] (D1a) at (-1,2) {$D_1$};
\node[draw=none] (D1b) at (1,2) {$D_1$};
\node[draw=none] (D1c) at (0,3) {$D_1$};
\node[draw=none] (f1a) at (-3,2) {$[2]_2$};
\node[draw=none] (f1b) at (3,2) {$[2]_2$};
\node[draw=none] (f1c) at (-3,3) {$[2]_2$};
\node[draw=none, left = of C1] (D1L1) {$D_{30}$};
\node[draw=none, left = of D1L1] (D1L2) {$C_{15}$};
\node[draw=none, left = of D1L2] (D1L3) {$D_1$};
\draw[solid] (C1)--(D10)--(D11);
\draw[solid,red,thick] (C1) --  (D1a) ;
\draw[solid,red,thick] (C1) -- node[midway,below right] {\scriptsize $3$} (D1b) ;
\draw[solid,red,thick] (C1) -- (D1c) ;
\draw[solid,gray,thick] (D1L1) -- node[midway,above left] {\scriptsize $2$} (D1a) ;
\draw[solid,gray,thick] (D1L1) -- (D1b) ;
\draw[solid,gray,thick] (D1L1) -- (D1c) ;
\draw[solid,blue,thick] (D1a)--(D1b) ;
\draw[solid,blue,thick] (D1a)--(D1c) ;
\draw[solid,blue,thick] (D1b)-- node[midway,above right] {\scriptsize $6$} (D1c) ;
\draw[solid] (D1L3)--(D1L2)--(D1L1)--(C1);
\draw[solid,thick, snake it] (D1a)--(f1a) ;
\draw[solid,thick, snake it] (D1b)--(f1b) ;
\draw[solid,thick, snake it] (D1c)--(f1c) ;
\end{tikzpicture}
}
\quad /\BZ_2}
There are $3$ free hypermultiplets in this example.
\ei

\subsubsection*{Consistency checks}
We can perform a similar consistency check of the proposed mirror theory as in precedent subsections. Let us demonstrate this in two examples:
\bi
\item {\bf Let us take $N=8$ and $p=10$}.  Here $x=1$. The 4d theory has $24(c-a) = 281/5 = 56+(1/5)$ and it has rank $35$.  The proposed mirror theory has CB dimension $56$ and HB dimension $34$.  We thus have $35-34=1$ free hypermultiplets, in agreement with the fact that the non-Higgsable SCFT has $24(c-a)$ equal to $1/5$, which is expected to be that of the $(A_1, A_2)$ theory whose rank is $1$.
\item {\bf Let us take $N=40$ and $p=30$}.  Here $x=2$.  The 4d theory has $24(c-a) = 7658/5 = 1531+(3/5)$ and it has rank $578$.  The proposed mirror theory has CB dimension $1531$ and HB dimension $575$.  We thus have $578-575=3$ free hypermultiplets, in agreement with the fact that the non-Higgsable SCFT has $24(c-a)$ equal to $3/5$, which is expected to be that of the $(A_1, A_2)^{\otimes 3}$ theory whose rank is $3$.
\ei

\section{The 3d mirror of $(D_n,D_m)$ theories}\label{sec:new3dmirror}

In this section, we discuss the 3d mirror of $(D_n,D_m)$ theories where, without loss of generality, we assume that $m\geq n$.  It is convenient to arrange the discussion according to the number of mass parameters of the $(D_n,D_m)$ theories.  In terms of 3d mirror theories, such parameter corresponds to the rank of topological symmetry that arises from $D_1$ gauge nodes as well as the balanced $C$-type gauge nodes. All information regarding CB dimension, central charges, and number of mass parameters for all $(D_n,D_m)$ theories can be computed, for instance, using the code given in~\cite{Carta:2020plx}.

\subsection{Theories with $1$ mass parameter}
\label{sec:DpDq1mass}
The $(D_{n}, D_{m})$ theories with one mass parameter have both $n$ and $m$ odd and either $\frac{n-1}{\GCD(n-1,m-1)}$ or $\frac{m-1}{\GCD(n-1,m-1)}$ is even; see \eref{masscount}. For definiteness, we take $m>n$.
According to \eref{eq:HBdimDnDm}, such a theory has Higgs branch dimension $(n-1)/2$.  Indeed, this is in agreement with the observation that the value of $24(c-a)$ of the 4d theory is at least $(n-1)/2$, with a possibility of an additional fractional number. The latter corresponds to the value of $24(c-a)$ of the non-Higgsable sector present on the Higgs branch of the 4d theory. On the other hand, the rank of the 4d theory in this class is $\frac{1}{2}(m n-1)$.  

Since the number of mass parameters of the 4d theory (which is one) corresponds to the rank of the topological symmetry of the 3d mirror theory, we conclude that the latter should contain either one $D_1$ gauge group or one balanced $C$-type gauge group, but not both. Given that the Higgs branch dimension of the 4d theory is $(n-1)/2$, the 3d mirror must have the Coulomb branch dimension $(n-1)/2$. A more plausible option would be the latter.   We thus propose that the mirror theory should be described by the $C_{(n-1)/2}$ SQCD with $n$ flavors, namely
\bes{ \label{CSQCD}
C_{(n-1)/2}-[D_n]~,
}
together with
\bes{ \label{Hfreeonemass}
H_{\text{free}} = \frac{1}{2}(mn-n^2+n-1) 
}
free hypermultiplets.  This number of free hypermultiplets precisely coincides with the rank of the non-Higgsable SCFT along the Higgs branch of the 4d theory.  In the next subsection, we provide a derivation of such a proposal for the $\left(D_{4\mathfrak{n}-1},D_{4\mathfrak{m}+4\mathfrak{n}-3}\right)$ theory, with $\fm, \fn \geq 1$.

Let us consider the special case of $n=3$ and $m=4 \fm+1$, \ie~  the $(D_3, D_{4 \fm+1}) \cong (A_3,D_{4\fm+1})$ theory. The 3d mirror is given below \cite[(6.7)]{Carta:2021whq}, with $\mu\rightarrow 2,\, \fm \rightarrow 1, \, \fN \rightarrow \fm$, and is described by the SQED with $4$ flavors with $6 \fm-2$ free hypermultiplets.  This theory is indeed dual to the $\USp(2)$ SQCD with $3$ flavors, given by \eref{CSQCD}, with $H_{\text{free}} = 6 \fm-2$ free hypermultiplets, given by \eref{Hfreeonemass}, as expected.

In the special case that $m-n$ divides $m-1$, we conjecture that the non-Higgsable SCFT can be identified as
\bes{ \label{nonHiggsableonemass}
\left(A_{m-n},D_{\frac{m-1}{m-n}+n-1}\right)~.
}
The value of $24(c-a)$ of this theory is precisely the difference between the value of $24(c-a)$ of the $(D_n,D_m)$ in question and $(n-1)/2$. Moreover, such a non-Higgsable SCFT has rank equal to $H_{\text{free}}$, given by \eref{Hfreeonemass}, as it should be. Let us demonstrate this in the example of $n=7$ and $m=9$, \ie~ the $(D_7, D_9)$ theory.  The value of $24(c-a)$ is $23/7 = 3+(2/7)$.  According to \eref{eq:HBdimDnDm}, the Higgs branch dimension of the $(D_7,D_9)$ theory is $3$.  Hence, the value of $24(c-a)$ of the non-Higgsable SCFT is $2/7$.  We identify the latter as the $(A_2, D_{10})$ theory, as claimed in \eref{nonHiggsableonemass}.  This theory has rank $10$, corresponding to $10$ free hypermultiplets as proposed in \eref{Hfreeonemass}.  

If $m-n$ does not divide $m-1$, the structure of non-Higgsable SCFT could be more complicated.  Let us consider the example of the $(D_5, D_{11})$ theory. The value of $24(c-a)$ is $2+(5/7)$ and the Higgs branch dimension is $2$, so we expect the non-Higgsable SCFT to have $24(c-a)$ equal to $5/7$. Since \eref{Hfreeonemass} gives $17$ free hypermultiplets, we expect also that the non-Higgsable SCFT has rank $17$.  One of the possibilities that fits these data is to identify the non-Higgsable SCFT in question as $(A_2, A_3) \otimes (A_2, D_4) \otimes (A_2, D_{10})$, although we cannot confirm this.  We leave the identification of the non-Higgsable SCFT in this case as an open problem for future work.

\subsubsection{Derivation of the 3d mirror for $\left(D_{4\mathfrak{n}-1},D_{4\mathfrak{m}+4\mathfrak{n}-3}\right)$}

In this section we focus on a subclass of theories with one mass parameter, \ie~ those with $n=4\fn-1$ and $m=4\fm-2+n$, namely $\left(D_{4\mathfrak{n}-1},D_{4\mathfrak{m}+4\mathfrak{n}-3}\right)$, with $\fm, \fn \geq 1$.

Following the notation of \eref{defalphabeta}, these theories correspond to $a=2, \, k=1,\, p=2\fn-1,\,  q=2\fm+2\fn-2$, and so
\begin{equation}
    \alpha=2\mathfrak{n}-1 \coma \beta=2\mathfrak{m}+2\mathfrak{n}-2~.
\end{equation}
Notice that $\beta-\alpha$ is odd and $\beta>\alpha$. According to \eref{eq:DnDmDecompCase1}, such a theory can be written as
\begin{equation} \label{SO4ngauge}
    \left(D_{4\mathfrak{n}-1},D_{4\mathfrak{m}+4\mathfrak{n}-3}\right)= D_{2\mathfrak{m}+4\mathfrak{n}-3}\left(\SO(4\mathfrak{m}+4\mathfrak{n}-2)\right)-\SO(4\mathfrak{n})-D_{2\mathfrak{m}+4\mathfrak{n}-3}\left(\SO(4\mathfrak{n})\right)\fstop
\end{equation}
where the full puncture of the theory on the left is partially closed to $[(2\fm-1)^2, 1^{4 \fn}]$. 

The 3d mirror of $D_{2\mathfrak{m}+4\mathfrak{n}-3}\left(\SO(4\mathfrak{m}+4\mathfrak{n}-2)\right)$ is described in Section \ref{sec:DpSOpodd} with $x=1$ and $X=\fm$, namely $T^{[2^{2 \fm},1^{4\fn-2}]}_{[1^{4\fm+4\fn-2}]} (\SO(4\fm+4\fn-2))$ with $2\fm(\fn-1)$ free hypermultiplets.  The partial closure of the puncture leads to the mirror theory $T^{[2^{2 \fm},1^{4\fn-2}]}_{[(2\fm-1)^2, 1^{4 \fn}]} (\SO(4\fm+4\fn-2))$, whose quiver is given by
\begin{equation}
\begin{array}{ccccccccccccccccc}
            &   &         &   &        &   &         &   &        &   &          &   &    [C_{\mathfrak{m}}]   \\
            &   &         &   &        &   &         &   &        &   &          &   &     |    \\
     D_1 & - & C_1 & - & D_2 & - & C_2 & - & \cdots & - & C_{2\mathfrak{n}-1} & - & D_{2\mathfrak{n}} & - & C_{2\mathfrak{n}-1} & - & [D_{2\mathfrak{n}-1}]
\end{array}
\end{equation}
together with $2\fm(\fn-1)$ free hypermultiplets.

On the other hand, the 3d mirror of $D_{2\mathfrak{m}+4\mathfrak{n}-3}\left(\SO(4\mathfrak{n})\right)$ is given by \cite[(5.4)]{Carta:2021whq}, with $\fm \rightarrow \fm$, $N \rightarrow 2 \fn$ and $\mu \rightarrow 1$.  Explicitly, this is the $T[\SO(4\fn))]$ theory, whose quiver is
\bes{
D_1-C_1-\cdots-D_{2\fn-1}-C_{2\fn-1}-[D_{2\fn}]~,
}
together with
\bes{
H_{\text{free}} = 2\fn(\fm-1)
}
free hypermultiplets.

We now gauge the common $\SO(4\fn)$ Coulomb branch symmetry of the two aforementioned mirror theories, as indicated in \eref{SO4ngauge}. As a result, we obtain
\bes{
[C_\fm]-[D_{2\fn}] - C_{2\fn-1}- [D_{2\fn-1}]
}
with $2\fm(\fn-1)+2\fn(\fm-1)$ free hypermultiplets.  This theory can be rewritten as
\bes{
C_{2\fn-1}- [D_{4\fn-1}]
}
with $2\fm(\fn-1)+2\fn(\fm-1)+4\fm \fn = 2(4 \fm \fn-\fm-\fn)$ free hypermultiplets.  This is in agreement as the proposal \eref{CSQCD} and \eref{Hfreeonemass}, as it should be.

\subsection{Theories with $2$ mass parameters}
According to \eref{masscount}, each $(D_n, D_m)$ theory with two mass parameters falls into one of the two categories: Either $n$ is even and $m$ is odd, or $n$ is odd and $m$ is even.  In the following discussion, we assume that $m>n$.

Let us first discuss the case of $n$ even and $m$ odd.  We propose that the mirror theory is
\bes{
T_{[(n-1)^2, 1^2]}[\SO(2n)]: \quad [D_n]-C_{n/2}-D_1
}
together with the following number of free hypermultiplets:
\bes{
H_{\text{free}}= \frac{1}{2} n( m-n-1)~.
}

On the other hand, for $n$ odd and $m$ even, we propose that the mirror theory is
\bes{
[C_F]-D_1-C_{(n-1)/2}-[D_{n-1}]
}
with
\bes{
F=\frac{1}{2}(m-n+1)~,
}
together with the following number of free hypermultiplets:
\bes{
H_{\text{free}} = (n-2)F =\frac{1}{2} (n-2)( m-n+1)~.
}

In each case, the two mass parameters of the 4d theory corresponds to two topological symmetries of the 3d mirror theory; one arises from the $D_1$ gauge group and the other arises from the balanced $C$-type gauge group. It can be checked that the Coulomb branch dimension of the mirror theory is in agreement with the Higgs branch of the 4d theory given by \eref{eq:HBdimDnDm}, and that the Higgs branch dimension of the mirror theory is in agreement with the rank of the 4d theory. Moreover, the difference between the value of $24(c-a)$ and the Higgs branch dimension of the 4d theory is in agreement with the value of $24(c-a)$ of the non-Higgsable SCFT, whose rank is in agreement with $H_{\text{free}}$ in each case.  As an example, we find that for the $(D_{2\fn+2}, D_{4\fn+3})$ theories, with $\fn\geq 1$, the non-Higgsable SCFT can be identified as $(A_{2\fn}, D_{2\fn+2})$, whose central charges satisfy $a=c$ and whose rank is equal to the number of free hypermultiplets, namely $2\fn(\fn+1)$, as expected.

\subsubsection{The special case of $(D_3, D_{2\mathfrak{n}+2})\cong (A_3, D_{2\mathfrak{n}+2})$ theories} 
\label{sec:D3twomasses}

This is a special case where $n=3$ and $m=2\fn+2$. In this case, from \eref{CSQCD} and \eref{Hfreeonemass}, the mirror theory of $(D_3, D_{2\fn+2}) \cong (A_3, D_{2\fn+2})$ is described by
\bes{
&[C_\fn]-D_1-C_1-[D_2] \\
&\text{with $H_{\text{free}} =\fn$ hypermultiplets}~.
}

However, as discussed in \cite{Carta:2021whq}, each theory in this class can be obtained from closing the maximal puncture of the $D^{4\fn+2}_{4\fn+6}(\SO(4\fn+4))$ theory. The mirror theory for $(D_3, D_{2\fn+2})=(A_3, D_{2\fn+2})$ admits the following two descriptions. One is discussed explicitly in \cite[Section 6.1.2]{Carta:2021whq}, namely
\bes{
&\begin{tikzpicture}[baseline]
\node[draw=none] at (0,0) {$D_1$}; 
\draw[solid, blue,thick] (0.3,0)--(1.5,0) node[above,midway] {$2$};
\draw[solid,snake it]
(2.1,0)--(3.1,0);
\node[draw=none] at (1.8,0) {$D_1$};
\node[draw=none] at (3.6,0) {$[2\fn]_2$};
\end{tikzpicture} \qquad /\BZ_2 \\
&\text{+ $\fn$ free hypermultiplets}
}
The other description is a simple modification of \cite[(4.17)]{Carta:2021whq}, namely
\bes{
&\begin{tikzpicture}[baseline]
\node[draw, circle] (c1) at (-1,0) {$1$};
\node[draw, circle] (c2) at (1,0) {$1$};
\node[draw, circle] (c3) at (0,1) {$1$};
\draw[solid,thick,red] (c1)--(c2) node[below,midway] {$2\fn$};
\draw[solid,thick,blue] (c1)--(c3) node[left,midway] {\footnotesize $2$};
\draw[solid,thick,blue] (c2)--(c3) node[right,midway] {\footnotesize $2$};
\end{tikzpicture}
 \\
&\text{+ $\fn$ free hypermultiplets}
}
The $\fn$ free hypermultiplets arise from the non-Higgsable SCFT
\bes{
(A_1, A_{2\fn})~.
}

We thus have an isomorphism between the following three theories:
\bes{[C_\fn]-D_1-C_1-[D_2] \label{frame1}}
\bes{\begin{tikzpicture}[baseline]
\node[draw=none] at (0,0) {$D_1$}; 
\draw[solid, blue,thick] (0.3,0)--(1.5,0) node[above,midway] {$2$};
\draw[solid,snake it]
(2.1,0)--(3.1,0);
\node[draw=none] at (1.8,0) {$D_1$};
\node[draw=none] at (3.6,0) {$[2\fn]_2$};
\end{tikzpicture} \qquad /\BZ_2}
\bes{\begin{tikzpicture}[baseline]
\node[draw, circle] (c1) at (-1,0) {$1$};
\node[draw, circle] (c2) at (1,0) {$1$};
\node[draw, circle] (c3) at (0,1) {$1$};
\draw[solid,thick,red] (c1)--(c2) node[below,midway] {$2\fn$};
\draw[solid,thick,blue] (c1)--(c3) node[left,midway] {\footnotesize $2$};
\draw[solid,thick,blue] (c2)--(c3) node[right,midway] {\footnotesize $2$};
\end{tikzpicture}}
It can be checked, for example using the Hilbert series,\footnote{In general, the Higgs branch symmetry of each of these three theories is $\U(2\fn) \times \SO(4)$, and the Coulomb branch symmetry of each theory is $\SO(2)^2$. In the duality frame \eref{frame1}, the adjoint representation of the $\U(2\fn)$ factor of the Higgs branch symmetry arises from the representation $[0,\ldots,0]\oplus[2,0,\ldots,0]\oplus[0,1,0,\ldots,0]$ of the $C_\fn$ flavour symmetry.  As an example, for $\fn=3$, the unrefined Coulomb branch Hilbert series of these three theories reads
\bes{
1 + 2 t^2 + 5 t^4 + 8 t^6 + 17 t^8 + 26 t^{10} + 41 t^{12} + \ldots~, \nn
}
where $2$ is the dimension of Coulomb branch symmetry $\SO(2)^2$.
The unrefined Higgs branch Hilbert series of these three theories reads
\bes{
1 + 42 t^2 + 48 t^3 + 676 t^4 + 1200 t^5 + 6888 t^6 + 13920 t^7 + 
 52048 t^8+\ldots~, \nn
}
where $42$ is the dimension of the Higgs branch symmetry $\U(6) \times \SO(4)$.  These computations can be performed as shown in \cite[Appendix A]{Carta:2021whq}. 
} that the Higgs branches and the Coulomb branches of the three theories agree with each other.

\subsubsection{Derivation of the 3d mirror of $(D_{11},D_{10\fm+6})$}
Let us consider the $(D_{11},D_{10\fm+6})$ theory, with $\fm\geq 1$. Such a theory admits the following decomposition:
\begin{equation}
    (D_{11},D_{10\mathfrak{m}+6})=D_{6\mathfrak{m}+9}(\SO(12\mathfrak{m}+8))-\SO(10)-D_{4\mathfrak{m}+6}(\SO(10))\coma
\end{equation}
where $D_{6\mathfrak{m}+9}(\SO(12\mathfrak{m}+8))$ has a partially closed regular puncture labeled by the partition $\left[(6\mathfrak{m}-1)^2,1^{10}\right]$.

The mirror of $D_{4\mathfrak{m}+6}(\SO(10))$ has been found in~\cite[(6.7)]{Carta:2021whq}, with $\mu=1$ and $\fN=2$, to be $T^{\rho}_{\sigma}[\SO(8\mathfrak{m}+20)]$ with
\begin{equation}
    \rho=\left[3^8,2^{4\mathfrak{m}-2}\right] \coma \sigma=\left[(4\mathfrak{m}+5)^2,1^{10}\right]\coma
\end{equation}
whose quiver description is
\begin{equation}
    \begin{array}{ccccccccccccccccc}
      D_1 & - & C_1 & - & D_2 & - & C_2 & - & D_3 & - & C_3 & - & D_4 & - & C_4 & - & D_1\\
      & & & & & & & & & & & & & & | & &  |\\
      & & & & & & & & & & & & & & [D_4] & & [C_{2\mathfrak{m}-1}]
    \end{array}
    \label{eq:DpSO8m20}
\end{equation}
together with $H_{\text{free}}=6\mathfrak{m}-5$ free hypermultiplets. 

The mirror of $D_{6\mathfrak{m}+9}(\SO(12\mathfrak{m}+8))$ has been found using the procedure in Section~\ref{sec:DpSOpodd} to be $T^{\sigma}[\SO(12\mathfrak{m}+8)]$ with $\sigma=\left[2^{6\mathfrak{m}-2},1^{12}\right]$ and $H_{\text{free}}=15\mathfrak{m}-5$ free hypermultiplets. Partially closing the full puncture to $[(6\fm-1)^2, 1^{10}]$ yields the mirror theory $T^{\left[2^{6\mathfrak{m}-2},1^{12}\right]}_{\rho'}[\SO(12\mathfrak{m}+8)]$ where $\rho'=\left[(6\mathfrak{m}-1)^2,1^{10}\right]$, whose quiver description is
\begin{equation} \label{D6mp9partiallyclosed}
    \begin{array}{ccccccccccccccccccc}
      D_1 & - & C_1 & - & D_2 & - & C_2 & - & D_3 & - & C_3 & - & D_4 & - & C_4 & - & D_5 & - & C_5 \\
      & & & & & & & & & & & & & & & & | & &  |\\
      & & & & & & & & & & & & & & & & [C_{3\mathfrak{m}-1}] & & [D_{6}]
    \end{array}
\end{equation}

Finally, the mirror of $(D_{11},D_{10\fm+6})$ is obtained gauging the common $\SO(10)$ Coulomb branch symmetry between~\eqref{eq:DpSO8m20} and \eref{D6mp9partiallyclosed}.  This can be done by fusing the tails of the two theories together and splitting the $D_5$ node in the latter into $D_1$ and $[D_4]$, we obtain the following result
\bes{
\begin{tikzpicture}[baseline]
\node[draw=none] (g1) at (3,0) {$C_5$};
\node[draw=none] (g2) at (1.5,1) {$D_1$};
\node[draw=none] (f0) at (1.5,-1) {$[D_4]$};
\node[draw=none] (f2) at (1.5,2) {$[C_{2\fm-1}]$};
\node[draw=none] (f1) at (3,-1) {$[D_6]$};
\node[draw=none] (f3) at (0,0) {$[C_{3\fm-1}]$};
\draw(g1)--(g2);
\draw (g1)--(f1);
\draw (g1)--(f0);
\draw (f0)--(f3);
\draw (g2)--(f3);
\draw (g2)--(f2);
\end{tikzpicture}
}
together with $(6\fm-5)+(15\fm-5)=21\fm-10$ free hypermultiplets.  This can be rewritten as\footnote{This quiver has 
\bes{
\dim_{\BH}~\text{CB} = 6~, \quad \dim_{\BH}~ \text{HB} = 10\fm+50+H_{\text{free}}= 55\fm+32~. \nn
}
They agree with the Higgs branch dimension and the rank of the 4d theory in question, respectively.
}
\bes{ \label{mirrD11D10mp6}
[C_{5\fm-2}]-D_1-C_5-[D_{10}]
}
with $H_{\text{free}} =(24\fm-8)+(21\fm-10) = 45\fm-18 = 9(5\fm-2)$ free hypermultiplets.

\subsection{Theories with $2M+1$ mass parameters, with $M\geq 1$}
In this section, we focus on the $(D_n, D_m)$ theories with $2M+1$ mass parameters, where $M\geq 1$.  They can be parametrized as
\bes{ \label{parametrisationoddmass}
&n= 4M \fn- (2M-1)~, \quad m= n+4M \fm \\
& \GCD(2\fn-1,2\fm)=1~.
}
where $\fn, \fm \geq 1$ and we allow $\fm$ and $\fn$ to also take the values $(\fm=0, \fn=1)$.  We first discuss the latter case and then proceed with the general case.

\subsubsection{The $(D_{2M+1}, D_{2M+1})$ theory}
Let us consider the special case $\fm=0$ and $\fn=1$, namely, the $(D_{2M+1}, D_{2M+1})$ theory. It has rank $M(2M+1)$.  The value of $24(c-a)$, which is also equal to the Higgs branch dimension, is $3M$.
There are $2M+1$ mass parameters and $2M-1$ marginal operators. The central charges are:
\begin{equation} \label{acDoddDodd}
a=\frac{1}{24}M(4M+1)(4M+5)~, \quad    c=\frac{1}{3}M(M+1)(2M+1) \fstop
\end{equation}
This theory can also be realized using \eref{eq:DnDmDecompCase5} with $a=2M, \, k=M, \, p=q=1$ and $\alpha=\beta=M$:
\begin{equation} \label{DoddDoddglue}
\begin{array}{ccccc}
                                        & & [\SO(2)] &   \\
                                        & & | &   \\
 (D_{2M+1},D_{2M+1}) = D_{2M}(\USp(2M)) &- &\USp(2M) &- &D_{2M}(\USp(2M))
\end{array}
\end{equation}
where $[\SO(2)]$ means a \textit{full hypermultiplet} in the fundamental of $\USp(2M)$. 

We, in fact, observe that the $(D_{2M+1},D_{2M+1})$ also admits the following description:
\bes{ \label{classSDoddDodd}
&\text{a class $\CS$ theory associated with the $A_{2M}$ twisted sphere} \\ 
& \text{with $2N$ minimal untwisted punctures (each labelled by $[2M,1]$),} \\ 
& \text{and 2 minimal twisted punctures (each labelled by $[2M]_t$)},
}
It can indeed be checked that the $(a,c)$ central charges\footnote{The contribution to the effective number of hypermultiplets and vector multiplets $(n_h, n_v)$ of the punctures $[2M,1]$ and $[2M]_t$ are respectively $\left( (2 M + 1)^2,~ (2 M + 2) (2 M) \right)$ and $\left( \frac{2}{3} M \left(4 M^2+9 M+5\right),\frac{1}{6} M \left(16 M^2+36 M+23\right) \right)$. Adding these together with the contribution of the $A_{2N}$ sphere $\left(-\frac{4}{3} (2 M + 1) [(2 M + 1)^2 - 1), -\frac{4}{3} (2 M + 1) [(2 M + 1)^2 - 1] - 
 2 M \right)$, we obtain $(n_h, n_v)=\left( \frac{2}{3} M \left(4 M^2+6 M+5\right),\frac{1}{3} M \left(8 M^2+12 M+1\right) \right)$ of \eref{classSDoddDodd}. Using the relation $(a,c)=\left(\frac{2n_v+n_h}{12}, \frac{5n_v+n_h}{24} \right)$, we obtain the central charges of \eref{classSDoddDodd} to be $(a,c) = \left( \frac{1}{24} M (4 M+1) (4 M+5),\frac{1}{3} M (M+1) (2 M+1) \right)$. This is indeed equal to those of the $(D_{2M+1}, D_{2M+1})$ theory; see \eref{acDoddDodd}.} of the two theories are equal.  This provides a non-trivial test for the following duality:
 \bes{
 \eref{DoddDoddglue} \,\, \longleftrightarrow \,\, \eref{classSDoddDodd}~.
 }
 
 It was pointed out in \cite[(6.16)]{Closset:2020afy} that, upon reduction to 3d, the $(D_{2M+1},D_{2M+1})$ theory also admits the following quiver description with mixed unitary and special unitary gauge groups:
\bes{\label{lagD2Np1D2Np1a}
    \begin{array}{ccccccccccccc}
                  &   &        &   &   \SU(1) \\
                  &   &        &   &   | \\
                  &   &        &   &   \U(M) \\
                  &   &        &   &   | \\%
            \ \SU(1) & - & \U(M) & - & \SU(2M) & - & \SU(2M-1) & - & \cdots & - & \SU(2) & - & \SU(1) \\
    \end{array}
}
Since the class $\CS$ description of the $(D_{2M+1},D_{2M+1})$ theory is known to be \eref{classSDoddDodd}, we can apply the prescription provided in \cite{Benini:2010uu, Beratto:2020wmn} to find the mirror theory, which can be described as follows:
\bes{ \label{mirrDoddDodd}
\begin{tikzpicture}[baseline=0]
\node[draw=none] (C) at (0,0) {$C_M$}; 
\node[draw=none] (flv0) at (0,1) {$[D_1]$}; 
\node[draw=none] (D1a) at (-2,-1) {$D_1$}; 
\node[draw=none] (flva) at (-2,-2) {$[1]$}; 
\node[draw=none] (D1b) at (-1,-1) {$D_1$}; 
\node[draw=none] (flvb) at (-1,-2) {$[1]$}; 
\node[draw=none] (D1c) at (0,-1) {\Large $\cdots$}; 
\node[draw=none] (D1d) at (1,-1) {$D_1$};
\node[draw=none] (flvd) at (1,-2) {$[1]$}; 
\node[draw=none] (D1e) at (2,-1) {$D_1$};
\node[draw=none] (flve) at (2,-2) {$[1]$}; 
\draw (C)--(flv0);
\draw (C)--(D1a)--(flva);
\draw (C)--(D1b)--(flvb);
\draw (C)--(D1d)--(flvd);
\draw (C)--(D1e)--(flve);
\draw [decorate,decoration={brace,amplitude=10pt,raise=4pt,mirror},yshift=0pt]
(-2.5,-2.5) -- (2.5,-2.5) node [black] at (0,-3.5) {\footnotesize
$2M$ legs};
\end{tikzpicture}
\qquad \text{or} \qquad
\begin{tikzpicture}[baseline=0]
\node[draw=none] (C) at (0,0) {$C_M$}; 
\node[draw=none] (flv-1) at (-1,1) {$B_0$}; 
\node[draw=none] (flv-2) at (1,1) {$B_0$}; 
\node[draw=none] (D1a) at (-2,-1) {$\U(1)$}; 
\node[draw=none] (flva) at (-2,-2) {$[1]$}; 
\node[draw=none] (D1b) at (-1,-1) {$\U(1)$}; 
\node[draw=none] (flvb) at (-1,-2) {$[1]$}; 
\node[draw=none] (D1c) at (0,-1) {\Large $\cdots$}; 
\node[draw=none] (D1d) at (1,-1) {$\U(1)$};
\node[draw=none] (flvd) at (1,-2) {$[1]$}; 
\node[draw=none] (D1e) at (2,-1) {$\U(1)$};
\node[draw=none] (flve) at (2,-2) {$[1]$}; 
\draw (flv-1)--(C)--(flv-2);
\draw (C)--(D1a)--(flva);
\draw (C)--(D1b)--(flvb);
\draw (C)--(D1d)--(flvd);
\draw (C)--(D1e)--(flve);
\draw [decorate,decoration={brace,amplitude=10pt,raise=4pt,mirror},yshift=0pt]
(-2.5,-2.5) -- (2.5,-2.5) node [black] at (0,-3.5) {\footnotesize
$2M$ legs};
\end{tikzpicture}
}
Note that the quivers on the left and on the right are equivalent.  Each component in the right quiver in \eref{mirrDoddDodd} comes from
\bes{
T_{[2M,1]}[\SU(2M+1)]: &\quad U(1)-[2M+1] \quad \text{or} \quad  [2M]-U(1)-[1]  \\
T_{[2M]}[\USp(2M)]: &\quad [C_M]-B_0
}
such that the $C_M$ symmetry from each theory is commonly gauged.   In conclusion, we have found a new mirror pair, namely
\bes{ \label{newmirrorpairDoddDodd}
\eref{lagD2Np1D2Np1a}\quad \overset{\text{mirror}}{\longleftrightarrow} \quad \eref{mirrDoddDodd}~.
}

Note that for $M=1$, \eref{newmirrorpairDoddDodd} is self-mirror.  However, due to \cite[Section 4.1]{Beratto:2020wmn} and \cite[(4.16) with $\fm=1$]{Carta:2021whq}, we have the following dual descriptions:
\bes{
\eref{mirrDoddDodd}_{M=1} \quad \longleftrightarrow \quad 
\scalebox{1}{
\begin{tikzpicture}[scale=0.9,baseline=0,font=\footnotesize]
\tikzstyle{every node}=[minimum size=0.5cm]
\node (c2) at (0,1) {$D_1$};
\node (f1) at (-1.2,0)  {$D_1$}; 
\node (f2) at (1.2,0)  {$D_1$}; 
\draw[very thick,light-gray] (c2)--(f1) node[midway,left]  {\footnotesize $~$};
\draw[very thick, light-gray] (c2)--(f2) node[midway,right]   {\footnotesize $~$};
\draw[very thick, light-gray] (f1)--(f2) node[midway,below]  {\footnotesize $~$};
\end{tikzpicture}} \quad /\BZ_2
\quad \longleftrightarrow
\quad
\scalebox{1}{
\begin{tikzpicture}[scale=0.9,baseline=0,font=\footnotesize]
\node[draw,circle,inner sep=3pt] (n) at (0,1) {$1$};
\node[draw,circle,inner sep=3pt] (w) at (-1,0)  {$1$}; 
\node[draw,circle,inner sep=3pt] (s) at (0,-1)  {$1$};
\node[draw,circle,inner sep=3pt] (e) at (1,0) {$1$};
\draw[solid] (n)--(w)--(s)--(e)--(n);
\draw[solid] (n)--(s);
\draw[solid] (w)--(e);
\end{tikzpicture}}
}

\subsubsection{General result: the 3d mirror for theories with $2M+1$ mass parameters}
Based on \eref{mirrDoddDodd}, we now propose a prescription to construct the 3d mirror for the $(D_n,D_m)$ theories with $2M+1$ mass parameters.  

Let us adopt the parametrization \eref{parametrisationoddmass}.  The quiver description for the mirror theory in question contains the balanced $C_{(n-1)/2}= C_{M(2\fn-1)}$ central gauge node connected to one flavor $[D_1]$ node and to $2M$ $D_1$ gauge nodes in the following way.
\ben
\item Connect the $C_{M(2\fn-1)}$ central node to each $D_1$ gauge node with a red line with multiplicity $2\fn-1$.
\item Connect each pair of $D_1$ gauge nodes by a blue edge with multiplicity $\fm(2\fn-1)$. These form a complete graph with $2M$ nodes such that each node is connected by a blue line.
\item Each $D_1$ gauge group in the complete graph has 
\bes{
F=\fm(\fn+1)+(2\fn-1)
}
hypermultiplets with charge 1 under the corresponding $\U(1)\cong D_1$ gauge group.
\item There are
\bes{
H_{\text{free}} = 2M(\fm-1)(\fn-1) 
}
free hypermultiplets.  We conjecture that the non-Higgsable SCFTs are
\bes{ \label{nonHiggsableoddmass}
(A_{\fm-1}, A_{2\fn-2})^{\otimes (2M)}~.
}
\een

The quiver that we just described has
\bes{
\dim_\BH \, \text{CB} &= M + 2 M \fn~, \\
\dim_\BH \, \text{HB} &= M (2 M - 5) - [4 M (M - 1)] \fm - [2 M (4 M - 3)] \fn \\
&+ [2 M (4 M - 1)] \fm \fn + (8 M^2) \fn^2 + H_{\text{free}}~.
}
These are indeed in agreement with the Higgs branch dimension and the rank of the 4d theory, respectively.  Moreover, the value of $24(c-a)$ of the aforementioned non-Higgsable SCFTs plus the above $\dim_\BH \text{CB}$ indeed gives the value of $24(c-a)$ of the corresponding 4d theory, as it should be.  The $2M+1$ mass parameters in 4d theory corresponds to the $2M$ FI parameters associated with $2M$ $D_1$ gauge groups, and one hidden FI parameter \cite{Kapustin:1998fa} associated with the balanced $C_{M(2\fn-1)}$ gauge group.

Let us now consider this quiver theory in various special cases. Some of these also serve as a non-trivial test of the above proposal.

\subsubsection{Example: $M=1$, \ie~ three mass parameters}
We parametrize the theory in this class as $(D_{4 \fn-1}, D_{4\fn-1+4\fm})$ with $\GCD(2\fn-1, 2\fm)=1$.  The proposed mirror theory is
\bes{\label{mirrDoddDoddthreemasses}
\begin{tikzpicture}[baseline=0,font=\footnotesize]
\node[draw=none] (C) at (0,0) {$C_{2\fn-1}$}; 
\node[draw=none] (flv0) at (0,1) {$[D_1]$}; 
\node[draw=none] (D1a) at (-1.5,-1) {$D_1$}; 
\node[draw=none] (flva) at (-3,-1) {$[F]$}; 
\node[draw=none] (D1b) at (1.5,-1) {$D_1$}; 
\node[draw=none] (flvb) at (3,-1) {$[F]$}; 
\node[draw=none]  at (0,-2) {with $F= \fm(\fn+1)+(2\fn-1)$}; 
\draw (C)--(flv0);
\draw[blue,thick] (D1a)--(D1b) node[midway,below] {$\fm(2\fn-1)$};
\draw[red,thick] (C)--(D1a) node[midway, left, above, xshift=-0.3cm] {$2\fn-1$};
\draw (D1a)--(flva);
\draw[red, thick] (C)--(D1b);
\draw (D1b)--(flvb);
\end{tikzpicture}
}
together with
\bes{
H_{\text{free}} = 2(\fm-1)(\fn-1)
}
free hypermultiplets.  We emphasize that $[F]$ denotes $F$ hypermultiplets of charge $1$ under the corresponding $\U(1) \cong D_1$ in the quiver.

The special case of $\fn=1$, namely the $(D_3, D_{4\fm+3})=(A_3,D_{4\fm+3}) $,  is particularly interesting. In this case, \eref{mirrDoddDoddthreemasses} reduces to 
\bes{\label{mirrD3Doddthreemasses}
\begin{tikzpicture}[baseline=0,font=\footnotesize]
\node[draw=none] (C) at (0,0) {$C_{1}$}; 
\node[draw=none] (flv0) at (0,1) {$[D_1]$}; 
\node[draw=none] (D1a) at (-1.5,-1) {$D_1$}; 
\node[draw=none] (flva) at (-3,-1) {$[2\fm+1]$}; 
\node[draw=none] (D1b) at (1.5,-1) {$D_1$}; 
\node[draw=none] (flvb) at (3,-1) {$[2\fm+1]$}; 
\draw (C)--(flv0);
\draw[blue,thick] (D1a)--(D1b) node[midway,below] {$\fm$};
\draw[red,thick] (C)--(D1a) node[midway, left, above, xshift=-0.3cm] {$1$};
\draw (D1a)--(flva);
\draw[red, thick] (C)--(D1b);
\draw (D1b)--(flvb);
\end{tikzpicture}
}
In fact, there is another description for the mirror theory for $(A_3,D_{4\fm+3})$, given by \cite[(6.49)]{Carta:2021whq}\footnote{There is a minor typo in \cite[(6.49)]{Carta:2021whq} (version 2).  The correction should be as follows: The blue edge with multiplicity $M=\fm(2\fN-1)$ should be in between two $D_1$ nodes attached to the wiggle lines, whereas the remaining edges should be gray with multiplicity $\fm$.} with $\fm$ and $\fN$ in that reference set to $1$ and $\fm+1$ respectively:
\bes{\label{mirrD3DoddthreemassesA}
\begin{tikzpicture}[baseline=0,font=\footnotesize]
\node[draw=none] (C) at (0,1) {$D_{1}$}; 
\node[draw=none] (D1a) at (-1.5,0) {$D_1$}; 
\node[draw=none] (flva) at (-3,0) {$[\fm]_2$}; 
\node[draw=none] (D1b) at (1.5,0) {$D_1$}; 
\node[draw=none] (flvb) at (3,0) {$[\fm]_2$}; 
\draw[blue,thick] (D1a)--(D1b) node[midway,below] {$2\fm+1$};
\draw[gray,thick] (C)--(D1a) node[midway, left, above, xshift=-0.3cm] {$1$};
\draw[snake it] (D1a)--(flva);
\draw[gray, thick] (C)--(D1b);
\draw[snake it] (D1b)--(flvb);
\end{tikzpicture} \qquad /\BZ_2
}
where $[\fm]_2$ attached to the wiggle line denote $\fm$ hypermultiplets with charge $2$ under the corresponding $\U(1)=D_1$ node.  Note that \eref{mirrD3Doddthreemasses} and \eref{mirrD3DoddthreemassesA} have the same Coulomb branch dimension and the same Higgs branch dimension.  The Higgs branch symmetry of each of these theories is $\U(2\fm+1)^2 \times \U(\fm)^2 \times \U(1)$, as it is manifest in each description.  The Coulomb branch symmetry of each theory is $\U(1)^3$; in \eref{mirrD3Doddthreemasses} this arises from the $\U(1)$ topological symmetry of each of the two $D_1$ nodes and the emergent $\U(1)$ topological symmetry of the balanced $C_1$ node \cite{Gaiotto:2008ak}, whereas in \eref{mirrD3DoddthreemassesA} such as Coulomb branch symmetry arises from the $\U(1)$ topological symmetry of each of the three $D_1$ nodes. It can also be checked that the Coulomb branch Hilbert series and the Higgs branch Hilbert series of these two theories are equal.\footnote{As an example, for $\fm=2$, the unrefined Higgs branch Hilbert series of \eref{mirrD3Doddthreemasses} and \eref{mirrD3DoddthreemassesA} are
\bes{
1 + 59 t^2 + 248 t^3 + 2070 t^4 + 10440 t^5 + 54650 t^6+\ldots~, \nn
}
where $59$ is the dimension of the Higgs branch symmetry $\U(5)^2 \times \U(2)^2 \times \U(1)$, and the Coulomb branch Hilbert series of both theories are
\bes{
1 + 3 t^2 + 8 t^4 + 16 t^6 + 29 t^8 + 47 t^{10}+\ldots~, \nn
}
where $3$ is the dimension of the Coulomb branch symmetry $\U(1)^3$. These computations can be performed as shown in \cite[Appendix A]{Carta:2021whq}.
}  We thus claim the duality:
\bes{
\eref{mirrD3Doddthreemasses} \quad \longleftrightarrow \quad \eref{mirrD3DoddthreemassesA}~.
}
This also provides a non-trivial check of the proposal \eref{mirrDoddDoddthreemasses}.

\subsubsection{Example: $M=2$, \ie~ five mass parameters}
Each theory in this class can be written as $(D_{8 \fn-3} , D_{8\fn-3 +8\fm})$ with $\GCD(2\fn-1, 2\fm) =1$.
The proposed mirror theory is
\bes{\label{mirrDoddDoddfivemasses}
\begin{tikzpicture}[baseline=0,font=\footnotesize]
\node[draw=none] (C) at (0,0) {$C_{4\fn-2}$}; 
\node[draw=none] (flv0) at (0,1) {$[D_1]$}; 
\node[draw=none] (D1a) at (-1.5,-1) {$D_1$}; 
\node[draw=none] (flva) at (-3,-1) {$[F]$}; 
\node[draw=none] (D1b) at (1.5,-1) {$D_1$}; 
\node[draw=none] (flvb) at (3,-1) {$[F]$}; 
\node[draw=none] (D1c) at (-1.5,-2) {$D_1$}; 
\node[draw=none] (flvc) at (-3,-2) {$[F]$}; 
\node[draw=none] (D1d) at (1.5,-2) {$D_1$}; 
\node[draw=none] (flvd) at (3,-2) {$[F]$}; 
\node[draw=none]  at (0,-3.2) {with $F= \fm(\fn+1)+(2\fn-1)$}; 
\draw (C)--(flv0);
\draw[blue,thick] (D1a)--(D1b);
\draw[blue,thick] (D1a)--(D1c);
\draw[blue,thick] (D1a)--(D1d);
\draw[blue,thick] (D1b)--(D1c);
\draw[blue,thick] (D1b)--(D1d);
\draw[blue,thick] (D1c)--(D1d) node[midway,below] {$\fm(2\fn-1)$};
\draw[red,thick] (C)--(D1a) node[midway, left, above, xshift=-0.3cm] {$2\fn-1$};
\draw[red, thick] (C)--(D1b);
\draw[red, thick] (C)--(D1c);
\draw[red, thick] (C)--(D1d);
\draw (D1a)--(flva);
\draw (D1b)--(flvb);
\draw (D1c)--(flvc);
\draw (D1d)--(flvd);
\end{tikzpicture}
}
together with
\bes{
H_{\text{free}} = 4(\fm-1)(\fn-1)
}
free hypermultiplets.

Let us provide a check for the number of free hypermultiplets via the example of $\fn=2$ and $\fm=4$, \ie~ the $(D_{13}, D_{45})$ theory. The value of $24(c-a)$ of this theory is $82/7=10+(12/7)$, where $10$ is the Higgs branch dimension of the 4d theory, according to \eref{eq:HBdimDnDm}, which is in agreement with the Coulomb branch dimension of the mirror theory. The fraction $12/7$ can be identified as the value of $24(c-a)$ of the non-Higgsable SCFT $(A_2, A_3)^{\otimes 4}\cong(A_3, A_2)^{\otimes 4}$, in agreement with the claim \eref{nonHiggsableoddmass}.  Since this theory has rank $12$, we expect to have $12$ free hypermultiplets, as stated above.

\subsubsection{Example: $M=3$, \ie~ seven mass parameters}
Each theory in this class can be written as $(D_{12 \fn-5} , D_{12\fn-5 +12\fm})$ with $\GCD(2\fn-1, 2\fm) =1$.
The proposed mirror theory is
\bes{\label{mirrDoddDoddsevenmasses}
\begin{tikzpicture}[baseline=0,font=\footnotesize]
\node[draw=none] (C) at (0,0) {$C_{6\fn-3}$}; 
\node[draw=none] (flv0) at (0,1) {$[D_1]$}; 
\node[draw=none] (D1a) at (-1.5,-1) {$D_1$}; 
\node[draw=none] (flva) at (-3,-1) {$[F]$}; 
\node[draw=none] (D1b) at (1.5,-1) {$D_1$}; 
\node[draw=none] (flvb) at (3,-1) {$[F]$}; 
\node[draw=none] (D1c) at (-1.5,-2) {$D_1$}; 
\node[draw=none] (flvc) at (-3,-2) {$[F]$}; 
\node[draw=none] (D1d) at (1.5,-2) {$D_1$}; 
\node[draw=none] (flvd) at (3,-2) {$[F]$}; 
\node[draw=none] (D1e) at (-1,-3) {$D_1$}; 
\node[draw=none] (flve) at (-2.5,-3) {$[F]$}; 
\node[draw=none] (D1f) at (1,-3) {$D_1$}; 
\node[draw=none] (flvf) at (2.5,-3) {$[F]$}; 
\node[draw=none]  at (0,-4) {with $F= \fm(\fn+1)+(2\fn-1)$}; 
\draw (C)--(flv0);
\draw[blue,thick] (D1a)--(D1b);
\draw[blue,thick] (D1a)--(D1c);
\draw[blue,thick] (D1a)--(D1d);
\draw[blue,thick] (D1a)--(D1e);
\draw[blue,thick] (D1a)--(D1f);
\draw[blue,thick] (D1b)--(D1c);
\draw[blue,thick] (D1b)--(D1d);
\draw[blue,thick] (D1b)--(D1e);
\draw[blue,thick] (D1b)--(D1f);
\draw[blue,thick] (D1c)--(D1d);
\draw[blue,thick] (D1c)--(D1e);
\draw[blue,thick] (D1c)--(D1f);
\draw[blue,thick] (D1d)--(D1e);
\draw[blue,thick] (D1d)--(D1f);
\draw[blue,thick] (D1e)--(D1f) node[midway,below] {$\fm(2\fn-1)$};
\draw[red,thick] (C)--(D1a) node[midway, left, above, xshift=-0.3cm] {$2\fn-1$};
\draw[red, thick] (C)--(D1b);
\draw[red, thick] (C)--(D1c);
\draw[red, thick] (C)--(D1d);
\draw[red, thick] (C)--(D1e);
\draw[red, thick] (C)--(D1f);
\draw (D1a)--(flva);
\draw (D1b)--(flvb);
\draw (D1c)--(flvc);
\draw (D1d)--(flvd);
\draw (D1e)--(flve);
\draw (D1f)--(flvf);
\end{tikzpicture}
}
together with
\bes{
H_{\text{free}} = 6(\fm-1)(\fn-1)
}
free hypermultiplets.

Let us provide a check for the number of free hypermultiplets via the example of $\fn=2$ and $\fm=2$, \ie~ the $(D_{19}, D_{43})$ theory. The value of $24(c-a)$ of this theory is $81/5=15+(6/5)$, where $15$ is the Higgs branch dimension of the 4d theory, according to \eref{eq:HBdimDnDm}, which is in agreement with the Coulomb branch dimension of the mirror theory. The fraction $6/5$ can be identified as the value of $24(c-a)$ of the non-Higgsable SCFT $(A_1, A_2)^{\otimes 6}$, in agreement with the claim \eref{nonHiggsableoddmass}.  Since this theory has rank $6$, we expect to have $6$ free hypermultiplets, as stated above.

\subsection{Theories with $2M+2$ mass parameters, with $M\geq 1$}
A theory in this class can be written as $(D_{n},D_{m})$, such that
\bes{
&n = (4M-2)\fn-(2M-2)~, \quad m = n +(4M-2)\fm~, \\  
&\GCD(2\fn-1, 2\fm) =1
}
where $\fn, \fm \geq 1$ and we allow $\fm$ and $\fn$ to also take the values $(\fm=0, \fn=1)$.  We first discuss the latter case and then proceed with the general case.

\subsubsection{The $(D_{2M}, D_{2M})$ theory}
Let us focus on the case of $\fm=0$ and $\fn=1$, \ie~ the $(D_{2M}, D_{2M})$ theory. As pointed out in \cite[(6.6)]{Closset:2020afy}, this theory admits the following Lagrangian description
\begin{equation} \label{lagD2ND2N}
    \begin{array}{ccccccccccccc}
                  &   &        &   &   \SU(1) \\
                  &   &        &   &   | \\
                  &   &        &   &   \SU(M) \\
                  &   &        &   &   | \\%
            \ \SU(1) & - & \SU(M) & - & \SU(2M-1) & - & \SU(2M-2) & - & \cdots & - & \SU(2) & - & \SU(1) \\
    \end{array}
\end{equation}
where each $\SU(1)$ should be treated as one flavor of hypermultiplets transforming under the fundamental representation of the node next to it.  It can be checked that the Coulomb branch spectrum, the rank, the Higgs branch dimension, and the $(a,c)$ central charges of this quiver theory\footnote{The central charges are
\begin{equation}
    a=\frac{1}{24}\left(16M^3-7M-5\right) \coma c=\frac{1}{6}\left(4M^3-M-1\right)\fstop
\end{equation}
The Coulomb branch dimension (\ie~ the rank) is $(2M+1)(M-1)$, while the Higgs branch dimension is $3M+1$.  There are $2M+2$ masses and $2M$ marginal operators.} match perfectly with those of the $(D_{2M}, D_{2M})$ theory.

According to Section \ref{sec:confmanDnDm}, with $a=2M-1$, $p=q=1$, $k=M-1$, $\alpha=M-1$, $\beta=M$, there is another description of the $(D_{2M},D_{2M})$ theory, namely that given by \eref{eq:DnDmDecompCase1}: 
\begin{equation} \label{weakD2ND2N}
     (D_{2M},D_{2M}) = D_{2M}(\SO(2M+2)) -\SO(2M)- D_{2M-2}(\SO(2M)) ~.
\end{equation}
We can use this description to find the mirror theory for the $(D_{2M},D_{2M})$ theory.  Recall that the mirror theory of $D_{2n-2}(\SO(2n))$ is described by \cite[Section 6.2]{Carta:2021whq}, namely
\bes{ \label{mirrD2nm2SO2n}
\scalebox{1}{
\begin{tikzpicture}[baseline=0,font=\footnotesize]
\tikzstyle{every node}=[minimum size=0.5cm]
\node (D1) at (-6,0) {$D_{1}$};
\node (C1) at (-4.5,0) {$C_{1}$};
\node (dots) at (-3,0) {\large $\cdots$};
\node (DN) at (-1.5,0) {$D_{n-1}$};
\node (CN) at (0,0) {$C_{n-1}$};
\node (f1) at (1.5,2)  {$D_1$}; 
\node (f2) at (1.5,1)  {$D_1$}; 
\node (f3) at (1.5,0)  {\large \bf $\vdots$};
\node (f4) at (1.5,-1)  {$D_1$}; 
\draw[solid] (D1)--(C1)--(dots)--(DN)--(CN);
\draw[solid] (CN)--(f1) node[midway,left]  {\footnotesize $~$};
\draw[solid] (CN)--(f2) node[midway,left]  {\footnotesize $~$};
\draw[solid] (CN)--(f4) node[midway,left]  {\footnotesize $~$};
\node[draw=none] (Z2) at (4,-1)  {$/\BZ_2$};
\draw [decorate,decoration={brace,amplitude=10pt,raise=4pt},yshift=0pt]
(2,2.2) -- (2,-1.2) node [black,midway,xshift=1.8cm] {
$n$ nodes};
\end{tikzpicture}}
}
Considering such a theory for $n=M$ and for $n=M+1$ and gauging the common Coulomb branch symmetry $\SO(2M)$, as indicated in \eref{weakD2ND2N}, by fusing the two quiver tails together, we obtain the following mirror theory for $(D_{2M}, D_{2M})$:
\bes{ \label{mirrD2MD2M}
\scalebox{1}{
\begin{tikzpicture}[baseline=0,font=\footnotesize]
\tikzstyle{every node}=[minimum size=0.5cm]
\node (CN) at (0,0) {$C_{M}$};
\node (f1) at (1.5,2)  {$D_1$}; 
\node (f2) at (1.5,1)  {$D_1$}; 
\node (f3) at (1.5,0)  {\large \bf $\vdots$};
\node (f4) at (1.5,-1)  {$D_1$}; 
\draw[solid] (CN)--(f1) node[midway,left]  {\footnotesize $~$};
\draw[solid] (CN)--(f2) node[midway,left]  {\footnotesize $~$};
\draw[solid] (CN)--(f4) node[midway,left]  {\footnotesize $~$};
\node[draw=none] (Z2) at (6,0)  {$/\BZ_2$};
\draw [decorate,decoration={brace,amplitude=10pt,raise=4pt},yshift=0pt]
(2,2.2) -- (2,-1.2) node [black,midway,xshift=1.8cm] {$2M+1$ nodes};
\end{tikzpicture}}
}
The $2M+2$ mass parameters of the 4d theory corresponds to one hidden FI parameter of the balanced $C_M$ gauge node \cite{Kapustin:1998fa} and $2M+1$ FI parameters associated with each $D_1$ gauge node.

Since each gauge node in the quiver \eref{lagD2ND2N} has zero beta-function, the reduction to 3d yields a 3d $\CN=4$ gauge theory with the same quiver description \cite{Closset:2020afy,Giacomelli:2020ryy}.  We have thus established a new mirror pair, namely
\bes{
\eref{lagD2ND2N}_{3d} \,\, \overset{\text{mirror}}{\longleftrightarrow} \,\, \eref{mirrD2MD2M}~.
}

\subsubsection{General result: the 3d mirror for theories with $2M+2$ mass parameters}

We now provide a prescription to construct the quiver description for the corresponding 3d mirror theory as follows.  The quiver in question contain the balanced $C_{n/2}$ central gauge node connected to $2M+1$ $D_1$ gauge nodes in the following way.
\ben
\item Choose any two of the $D_1$ nodes (call them $B_1$ and $B_2$).  Connect each of them to the central $C_{n/2}$ node with a black line with multiplicity $1$.
\item Connect each of the rest of the $D_1$ nodes (call them $A_1$, $A_2$, $\cdots$, $A_{2M-1}$) to the central $C_{n/2}$ node with a red line with multiplicity $2\fn-1$.
\item Connect any two $A_i$ and $A_j$ nodes with a blue line with multiplicity $\fm(2\fn-1)$. These form a complete graph with $2M-1$ nodes such that each node is connected by a blue line.
\item Each of the $A_i$ nodes has $\fm(\fn - 1)$
flavors of hypermultiplets with carrying 2 under $\U(1) =D_1$. Each of the $B_i$ nodes has no flavor charged under it.
\item Connect the node $B_1$ to each of the $A_1$, $A_2$, $\cdots$, $A_{2M-1}$ nodes by a gray line with multiplicity $\fm$. 
\item Quotient the above theory by an overall $\BZ_2$ symmetry.
\item There are
\bes{
H_{\text{free}} = (2M-1)(\fm-1)(\fn-1)
}
free hypermultiplets. We conjecture that the non-Higgsable SCFTs are
\bes{ \label{nonHiggsableevenmass}
(A_{\fm-1}, A_{2\fn-2})^{\otimes (2M-1)}~.
}
\een
Note that the above procedure is very similar to that described in \cite[Section 6.1.2]{Carta:2021whq}. The quiver that we just described has
\begin{equation}
    \begin{split}
        \dim_\BH \text{CB} =&\, \fn(2M-1)+M+2\coma\\
        \dim_\BH \text{HB} =&\, \fm (2 M-1) [(M+1) (4 \fn-2)-7 \fn+5]+2 (M+1)^2 (1-2 \fn)^2 \\
        & \quad +(M+1) [6 (5-4 \fn) \fn-11]+9 \fn (2 \fn-3)+11 + H_{\text{free}} \\
        =&\, 2 \fm (2 M-3) [(M+1) (2 \fn-1)-3 \fn+2]+2 (M+1)^2 (1-2 \fn)^2 \\
        &\quad +(M+1) [4 (7-6 \fn) \fn-9]+2 (2-3 \fn)^2\fstop
    \end{split}
\end{equation}
These quantities are indeed in agreement with the Higgs branch dimension and the rank of the 4d theory, respectively.  Moreover, the value of $24(c-a)$ of the aforementioned non-Higgsable SCFTs plus the above $\dim_\BH \text{CB}$ indeed gives the value of $24(c-a)$ of the corresponding 4d theory, as it should be.  The $2M+2$ mass parameters in 4d theory corresponds to the $2M+1$ FI parameters associated with $2M+1$ $D_1$ gauge groups, and one hidden FI parameter associated with the balanced $C_{n/2}$ gauge group.

Let us now consider this quiver theory in various special cases. Some of these also serve as a non-trivial test of the above proposal.

\subsubsection{Example: $M=1$, \ie~ four mass parameters}
The mirror theory for $(D_{2\fn}, D_{2\fn+2\fm})$ with $\GCD(2\fn-1, 2\fm)=1$ is
\bes{ \label{mirr4masses}
&\begin{tikzpicture}[baseline]
\node[draw=none] (C) at (0,0) {$C_\fn$};
\node[draw=none] (B1) at (-2,0) {$D_1$};
\node[draw=none] (A1) at (0,-1.5) {$D_1$};
\node[draw=none] (B2) at (2,0) {$D_1$};
\node[draw=none] (F1) at (0,-2.9) {$[\fm(\fn-1)]_2$};
\node[draw=none] (Z2) at (3.5,-1.5) {$/\BZ_2$};
\draw[solid, thick, black] (C)--(B1);
\draw[solid, thick, black] (C)--node[midway, above]{\scriptsize $1$}(B2);
\draw[solid, thick, red] (C)--node[midway,right, yshift=-0.3cm]{\scriptsize $2\fn-1$} (A1);
\draw[thick, snake it] (A1)--(F1);
\draw[thick, gray] (B1)--node[midway,below]{\scriptsize $\fm$}(A1);
\end{tikzpicture}\\
&\text{+ $(\fm-1)(\fn-1)$ free hypermultiplets}
}
This theory has
\bes{
\dim_\BH \text{HB}\, \eref{mirr4masses}  &=  \fm \fn+\fm+2 \fn^2+\fn-3 +H_{\text{free}} = 2 \fn (\fm+\fn)-2 \\
\dim_\BH \text{CB}\, \eref{mirr4masses}  &= \fn+3~.
}

Note that the special case of $\fn=1$ corresponds to $(D_2, D_{2+2\fm})=D_{2\fm+2}(\SO(4))$; the  theory \eref{mirr4masses} reduces to the mirror theory of $D_{2\fm+2}(\SO(4))$ described by \cite[(6.34)]{Carta:2021whq} with $\fN=1$.

\subsubsection{Example: $M=2$, \ie~ six mass parameters}
The mirror theory for $(D_{6\mathfrak{n}-2},D_{6\mathfrak{n}+6\mathfrak{m}-2})$ with $\GCD(2\fn-1, 2\fm)=1$ is
\bes{ \label{mirr6masses}
&\scalebox{0.9}{
\begin{tikzpicture}[baseline]
\node[draw=none] (C) at (0,1.2) {$C_{3\fn-1}$};
\node[draw=none] (B1) at (-3,1.2) {$D_1$};
\node[draw=none] (A1) at (-2,-1) {$D_1$};
\node[draw=none] (A2) at (0,-2) {$D_1$};
\node[draw=none] (A3) at (2,-1) {$D_1$};
\node[draw=none] (B2) at (3,1.2) {$D_1$};
\node[draw=none] (F1) at (-2,-2.4) {$[\fm(\fn-1)]_2$};
\node[draw=none] (F2) at (0,-3.3) {$[\fm(\fn-1)]_2$};
\node[draw=none] (F3) at (2,-2.4) {$[\fm(\fn-1)]_2$};
\node[draw=none] (Z2) at (3.5,-1.5) {$/\BZ_2$};
\draw[solid, thick, black] (C)--(B1);
\draw[solid, thick, black] (C)--node[midway, above]{\scriptsize $1$}(B2);
\draw[solid, thick, red] (C)--(A1);
\draw[solid, thick, red] (C)--(A2);
\draw[solid, thick, red] (C)--node[midway,right]{\scriptsize $2\fn-1$}(A3);
\draw[thick, snake it] (A1)--(F1);
\draw[thick, snake it] (A2)--(F2);
\draw[thick, snake it] (A3)--(F3);
\draw[thick, blue] (A1)--(A2)--(A3)-- node[midway,above]{\scriptsize $\fm(2\fn-1)$}(A1);
\draw[thick, blue] (A1)--(A3);
\draw[thick, gray] (B1)--node[midway,left]{\scriptsize $\fm$}(A1);
\draw[thick, gray] (B1)--(A2);
\draw[thick, gray] (B1)--(A3);
\end{tikzpicture}}\\
&\text{+ $3(\fm-1)(\fn-1)$ free hypermultiplets}
}
This theory has
\bes{
\dim_\BH \text{HB}\, \eref{mirr6masses}  &=  3 \fm (5 \fn-1)+9 \fn (2 \fn-1)-4 +H_{\text{free}} \\
&= 6 \fm (3 \fn-1)+6 \fn (3 \fn-2)-1 \\
\dim_\BH \text{CB}\, \eref{mirr6masses}  &= 3\fn+4~.
}

In Section \ref{sec:derive6masses}, we shall derive the mirror theory \eref{mirr6masses} for a subclass of theories with $\fn=\fN$ and $\fm=1$, \ie~ $(D_{6\fN-2}, D_{6\fN+4})$.  This serves as a highly non-trivial test of the proposal \eref{mirr6masses}.

Let us also provide a check for the number of free hypermultiplets via the example of $\fn=2$ and $\fm=4$, \ie~ the $(D_{10}, D_{34})$ theory. The value of $24(c-a)$ of this theory is $79/7=10+(9/7)$, where $10$ is the Higgs branch dimension of the 4d theory, according to the last case of \eref{eq:HBdimDnDm}, which is in agreement with the Coulomb branch dimension of the mirror theory. The fraction $9/7$ can be identified as the value of $24(c-a)$ of the non-Higgsable SCFT $(A_2, A_3)^{\otimes 3}\cong(A_3, A_2)^{\otimes 3}$, in agreement with the claim \eref{nonHiggsableevenmass}.  Since this theory has rank $9$, we expect to have $9$ free hypermultiplets, as stated above.

\subsubsection{Example: $M=3$, \ie~ eight mass parameters}
The mirror theory for $(D_{10\mathfrak{n}-4},D_{10\mathfrak{n}+10\mathfrak{m}-4})$ with $\GCD(2\fn-1, 2\fm)=1$ is
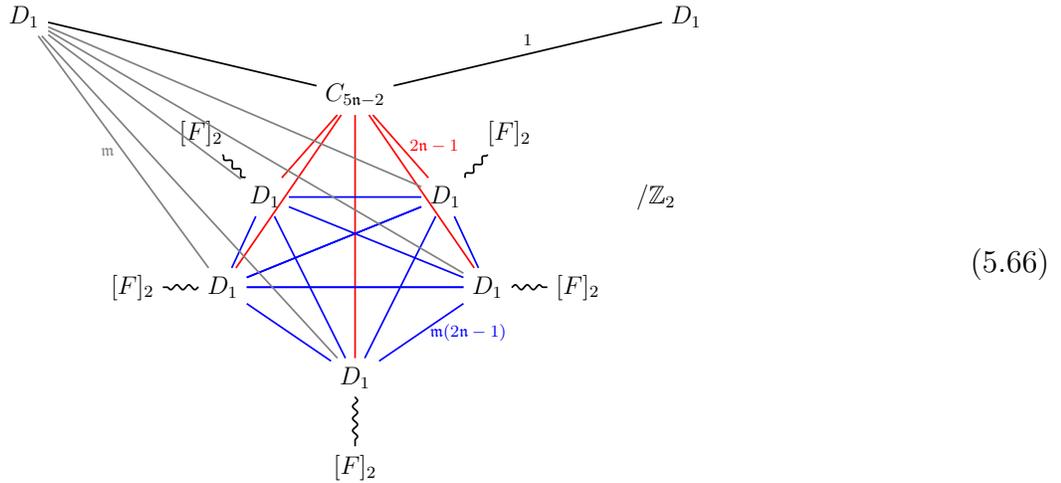
\bes{ \label{mirr8masses}
& \scalebox{0.8}{
\begin{tikzpicture}[baseline]
\node[draw=none] (C) at (0,0.2) {$C_{5\fn-2}$};
\node[draw=none] (B1) at (-5.5,1.5) {$D_1$};
\node[draw=none] (A1) at (-2.2,-3) {$D_1$};
\node[draw=none] (A2) at (0,-4.5) {$D_1$};
\node[draw=none] (A3) at (2.2,-3) {$D_1$};
\node[draw=none] (A4) at (-1.5,-1.5) {$D_1$};
\node[draw=none] (A5) at (1.5,-1.5) {$D_1$};
\node[draw=none] (B2) at (5.5,1.5) {$D_1$};
\node[draw=none, left of = A1, node distance=1.5cm] (F1) {$[F]_2$};
\node[draw=none, below of = A2, node distance=1.5cm] (F2) {$[F]_2$};
\node[draw=none, right of = A3, node distance=1.5cm] (F3) {$[F]_2$};
\node[draw=none, above left of = A4, node distance=1.5cm] (F4) {$[F]_2$};
\node[draw=none, above right of = A5, node distance=1.5cm] (F5) {$[F]_2$};
\node[draw=none] (Z2) at (5,-1.5) {$/\BZ_2$};
\draw[solid, thick, black] (C)--(B1);
\draw[solid, thick, black] (C)--node[midway, above]{\scriptsize $1$}(B2);
\draw[solid, thick, red] (C)--(A1);
\draw[solid, thick, red] (C)--(A2);
\draw[solid, thick, red] (C)--(A3);
\draw[solid, thick, red] (C)--(A4);
\draw[solid, thick, red] (C)--node[midway,right]{\scriptsize $2\fn-1$}(A5);
\draw[thick, snake it] (A1)--(F1);
\draw[thick, snake it] (A2)--(F2);
\draw[thick, snake it] (A3)--(F3);
\draw[thick, snake it] (A4)--(F4);
\draw[thick, snake it] (A5)--(F5);
\draw[thick, blue] (A1)--(A2)--node[midway,right]{\scriptsize $\fm(2\fn-1)$}(A3)--(A4)--(A5)--(A1);
\draw[thick, blue] (A3)--(A1);
\draw[thick, blue] (A1)--(A3);
\draw[thick, blue] (A1)--(A4);
\draw[thick, blue] (A1)--(A5);
\draw[thick, blue] (A2)--(A4);
\draw[thick, blue] (A2)--(A5);
\draw[thick, blue] (A3)--(A5);
\draw[thick, gray] (B1)--node[midway,left]{\scriptsize $\fm$}(A1);
\draw[thick, gray] (B1)--(A4);
\draw[thick, gray] (B1)--(A5);
\draw[thick, gray] (B1)--(A3);
\draw[thick, gray] (B1)--(A2);
\end{tikzpicture}}\\
&\text{+ $5(\fm-1)(\fn-1)$ free hypermultiplets, and with $F= \fm(\fn-1)$}
}
This theory has
\bes{
\dim_\BH \text{HB}\, \eref{mirr8masses}  &=  15 \fm (3 \fn-1)+5 \fn (10 \fn-7)-1 +H_{\text{free}} \\
&= 10 \fm (5 \fn-2)+10 \fn (5 \fn-4)+4 \\
\dim_\BH \text{CB}\, \eref{mirr8masses}  &= 5\fn+5~.
}

Let us also provide a check for the number of free hypermultiplets via the example of $\fn=2$ and $\fm=4$, \ie~ the $(D_{16}, D_{56})$ theory. The value of $24(c-a)$ of this theory is $120/7=15+(15/7)$, where $15$ is the Higgs branch dimension of the 4d theory, according to \eref{eq:HBdimDnDm}, which is in agreement with the Coulomb branch dimension of the mirror theory. The fraction $15/7$ can be identified as the value of $24(c-a)$ of the non-Higgsable SCFT $(A_2, A_3)^{\otimes 5}\cong(A_3, A_2)^{\otimes 5}$, in agreement with the claim \eref{nonHiggsableevenmass}.  Since this theory has rank $15$, we expect to have $15$ free hypermultiplets, as stated above.

\subsubsection{Derivation of \eref{mirr6masses} in a special case of $(D_{6\fN-2}, D_{6\fN+4})$} \label{sec:derive6masses}
Let use take $\fn=\fN$ and $\fm=1$, \ie~ $(D_{6\fN-2}, D_{6\fN+4})$.
This theory admits two descriptions.  One description corresponds to \eref{eq:DnDmDecompCase1} with $(\alpha=2\fN-1, \beta=4\fN+2)$:
\bes{ \label{cusp1a}
(D_{6\fN-2}, D_{6\fN+4}) = D_{8 \fN}(\SO(8\fN+6)) - \SO(4\fN) - D_{4 \fN}(\SO(4\fN))~,
}
with the left matter sector has a full puncture partially closed to $[(2\fN+3)^2, 1^{4 \fN}]$.  The other description corresponds to \eref{eq:DnDmDecompCase2} with $(\alpha=4\fN-2, \beta=2\fN+1)$:
\bes{ \label{cusp2a}
(D_{6\fN-2}, D_{6\fN+4}) = D_{4\fN}(\SO(4\fN+4)) - \SO(4\fN+4) - D_{8\fN}(\SO(8\fN-2))~,
}
with the right matter sector has a full puncture partially closed to $[(2\fN-3)^2, 1^{4\fN+4}]$.

For this theory, the mirror theory is given by \eref{mirr6masses} with $\fn=\fN$ and $\fm=1$. We redraw it as follows.
\bes{ \label{eq:6massesDnDmwithm=1}
&\begin{tikzpicture}[scale=1.2,baseline]
\node[draw=none] (g1) at (2.2,0) {\footnotesize $C_{3\fN-1}$};
\node[draw=none] (g2) at (1.2,1) {$D_1$};
\node[draw=none] (g3) at (1.2,-1) {$D_1$};
\node[draw=none] (g4) at (5.5,0) {$D_1$};
\node[draw=none] (g5) at (4,1) {$D_1$};
\node[draw=none] (g6) at (4,-1) {$D_1$};
\node[draw=none] (f1) at (1.2,2.2) {$[\fN-1]_2$};
\node[draw=none] (f2) at (4,2.2) {$[\fN-1]_2$};
\node[draw=none] (f3) at (7,0) {$[\fN-1]_2$};
\node[draw=none] (Z2) at (10,0) {$/\BZ_2$};
\draw[thick,red](g1)--  (g2);
\draw[thick,red](g1)-- node[below right ,midway]{\scriptsize{$2\fN-1$}} (g4);
\draw[thick,red](g1)--  (g5);
\draw[very thick](g1)--  (g3);
\draw[thick](g1)--  (g6);
\draw[thick,gray](g6)--(g2);
\draw[thick,gray](g5)--(g6);
\draw[thick,gray](g6)--node[below right ,midway]{\scriptsize{$1$}} (g4);
\draw[thick,blue](g5) --node[above right,midway,xshift=-0.4cm]{\footnotesize$2\fN-1$} (g4);
\draw[thick,draw,solid,black,snake it] (g2) -- (f1);
\draw[thick,draw,solid,black,snake it] (g5) --(f2);
\draw[thick,draw,solid,black,snake it] (g4) -- (f3);
\draw[thick,blue](g2)--(g5);
\draw[thick,blue](g2)--(g4);
\end{tikzpicture}
}
There are no free hypermultiplets for this mirror theory. 

We now derive \eref{eq:6massesDnDmwithm=1} by gluing the mirror theories of those indicated in \eref{cusp1a} and \eref{cusp2a}.


\subsubsection*{Gluing via \eref{cusp1a}}
Let us first consider \eref{cusp1a}.  The mirror theory for $D_{8 \fN}(\SO(8\fN+6))$ is given in Section \ref{sec:DpSOpeven-ratioodd}, with $N=4\fN+3$, $p=8\fN$, $2N-2=8\fN+4$, $\GCD(2N-2,p)=4$, $x=1$:
\bes{
\scalebox{0.85}{
\begin{tikzpicture}[baseline]
\node[draw=none] (C1) at (0,0) {$C_{4\fN}$};
\node[draw=none, right = of C1] (D10)  {$D_{4\fN+2}$};
\node[draw=none, right = of D10] (D11) {$C_{4\fN+1}-D_{4\fN+1}-\cdots-C_1-D_1$};
\node[draw=none] (D1a) at (-1,2) {$D_1$};
\node[draw=none] (D1b) at (1,2) {$D_1$};
\node[draw=none] (f1a) at (-3,2) {$[\fN-1]_2$};
\node[draw=none] (f1b) at (3,2) {$[\fN-1]_2$};
\node[draw=none, left = of C1] (D1L1) {$D_{1}$};
\draw[solid] (C1)--(D10)--(D11);
\draw[solid,red,thick] (C1) -- node[midway,below left] {\scriptsize $2\fN-1$} (D1a) ;
\draw[solid,red,thick] (C1) -- (D1b) ;
\draw[solid,blue,thick] (D1a)-- node[midway,above] {\scriptsize $2\fN-1$} (D1b) ;
\draw[solid, very thick] (D1L1)--(C1);
\draw[solid, thick, gray] (D10) -- (D1a);
\draw[solid, thick, gray] (D10) -- node[midway,above right] {\scriptsize $1$} (D1b);
\draw[solid,thick,snake it] (D1a) -- (f1a) ;
\draw[solid,thick,snake it] (D1b) -- (f1b) ;
\end{tikzpicture}
}
}
Upon partially closing the full puncture of $D_{8\fN}(\SO(8\fN+6))$ to $[(2\fN+3)^2, 1^{4\fN}]$, the mirror theory becomes
\bes{ \label{glue1a}
\scalebox{0.85}{
\begin{tikzpicture}[baseline]
\node[draw=none] (C1) at (0,0) {$C_{3\fN-1}$};
\node[draw=none, right = of C1] (D10)  {$D_{2\fN}$};
\node[draw=none, right = of D10] (D11) {$C_{2\fN-1}-D_{2\fN-1}-\cdots-C_1-D_1$};
\node[draw=none] (D1a) at (-1,2) {$D_1$};
\node[draw=none] (D1b) at (1,2) {$D_1$};
\node[draw=none] (f1a) at (-3,2) {$[\fN-1]_2$};
\node[draw=none] (f1b) at (3,2) {$[\fN-1]_2$};
\node[draw=none, left = of C1] (D1L1) {$D_{1}$};
\draw[solid] (C1)--(D10)--(D11);
\draw[solid,red,thick] (C1) -- node[midway,below left] {\scriptsize $2\fN-1$} (D1a) ;
\draw[solid,red,thick] (C1) -- (D1b) ;
\draw[solid,blue,thick] (D1a)-- node[midway,above] {\scriptsize $2\fN-1$} (D1b) ;
\draw[solid, very thick] (D1L1)--(C1);
\draw[solid, thick, gray] (D10) -- (D1a);
\draw[solid, thick, gray] (D10) -- node[midway,above right] {\scriptsize $1$} (D1b);
\draw[solid,thick,snake it] (D1a) -- (f1a) ;
\draw[solid,thick,snake it] (D1b) -- (f1b) ;
\end{tikzpicture}
}
}
where the tail $D_1-C_1-\cdots-D_{2\fN-1}-C_{2\fN-1}$ gives rise to the $\SO(4\fN)$ CB symmetry corresponding to the part $1^{4\fN}$ of the partition, whereas the balanced gauge group $C_{3\fN-1}$ gives rise to the $\SO(2)$ CB symmetry corresponding to the part $(2\fN-3)^2$ of the partition.  Note that the $D_{2\fN}$ gauge node is overbalanced.  Indeed, the tail on the right is determined by the $T_{[(2\fN+3)^2, 1^{4\fN}]}(\SO(8\fN+6))$ theory.  

On the other hand, the mirror theory for $D_{4 \fN}(\SO(4\fN))$ is given by \cite[(6.35)]{Carta:2021whq} with $\fm=1$:
\bes{ \label{glue2a}
\begin{tikzpicture}[baseline]
\node[draw=none] (D10) at (6,0) {$D_{2 \fN-2}-C_{2\fN-3}-D_{2\fN-3}-\cdots-C_1-D_1$};
\node[draw=none] (C1) at (1,0) {$C_{2\fN-1}$};
\node[draw=none] (F) at (-3,1) {$[\fN-1]_2$};
\node[draw=none] (D1a) at (-1,1) {$D_1$};
\node[draw=none] (D1b) at (-1,-1){$D_1$};
\draw[solid, thick, snake it] (F) to (D1a);
\draw[solid,gray,thick] (D1a) to (D1b) node at (-1.8,0) {$1$};
\draw[solid] (C1) to (D10);
\draw[solid,red,thick] (C1) -- node[midway,above right] {\scriptsize $2\fN-1$} (D1a) ;
\draw[solid, thick] (C1) to (D1b);
\end{tikzpicture}
}

Now we glue \eref{glue1a} and \eref{glue2a} together by fusing the tails $D_1-C_1-\cdots-D_{2\fN-1}-C_{2\fN-1}$ of the two quivers.  The latter corresponds to commonly gauging the CB symmetry $\SO(4\fN)$ of the two theories, as instructed by \eref{cusp1a}. In doing so, we split the $D_{2\fN}$ into two $D_1$ nodes, with the connections as depicted in the left most part of \eref{glue2a}. These two $D_1$ nodes become those linked by the gray line with label ``$1$'' in \eref{glue1a}.  The leftmost $D_1$ node in \eref{glue1a} becomes the lower left $D_1$ node in \eref{eq:6massesDnDmwithm=1}. This part of gluing indeed explains the $D_1-C_{3\fn-1}$ tail in \eref{mirr6masses}. The top two $D_1$ nodes linked by the blue line in \eref{glue1a} become the top two $D_1$ nodes in \eref{eq:6massesDnDmwithm=1}. Each of the three $D_1$ nodes attached to the wiggle line are connected together by the blue line, as in \eref{glue1a}. We thus arrive at \eref{eq:6massesDnDmwithm=1} as expected.

\subsubsection*{Gluing via \eref{cusp2a}}
The mirror theory for $D_{8\fN}(\SO(8\fN-2))$ is given by \cite[(6.48)]{Carta:2021whq} with $\fm=1$:
\bes{
\begin{tikzpicture}[scale=1.2,baseline]
\node[draw=none] (g0) at (0,0) {$D_1-C_1-\cdots-D_{4\fN-2}$};
\node[draw=none] (g1) at (2.8,0) {$C_{4\fN-2}$};
\node[draw=none] (g4) at (6,0) {$D_1$};
\node[draw=none] (g5) at (4.5,1) {$D_1$};
\node[draw=none] (g6) at (4.5,-1) {$D_1$};
\node[draw=none] (f2) at (4.5,2) {$[\fN-1]_2$};
\node[draw=none] (f3) at (7.5,0) {$[\fN-1]_2$};
\draw[black](g0)--(g1);
\draw[thick,red](g1)--(g4);
\draw[thick,red](g1) -- node[midway,above left,xshift=0.2cm] {\scriptsize $2\fN-1$}  (g5);
\draw[thick](g1)--(g6);
\draw[thick,gray](g5)--(g6);
\draw[thick,gray](g6)-- node[below right,midway]{\scriptsize $1$} (g4);
\draw[thick,blue](g5) --node[above right,midway,xshift=-0.4cm]{\scriptsize$2\fN-1$} (g4);
\draw[thick,draw,solid,black,snake it] (g5) --(f2);
\draw[thick,draw,solid,black,snake it] (g4) -- (f3);
\end{tikzpicture}
}
Upon partially closing the full puncture of $D_{8\fN}(\SO(8\fN-2))$ to $[(2\fN-3)^2, 1^{4\fN+4}]$, the mirror theory becomes
\bes{ \label{mirrD8NSO8Nm2}
\begin{tikzpicture}[scale=1,baseline, font=\footnotesize]
\node[draw=none] (g1) at (2.8,0) {$C_{3\fN}$};
\node[draw=none, left = of g1] (g0) {$D_{2\fN+2}$};
\node[draw=none, left = of g0] (gm1)  {$D_1-C_1-\cdots-D_{2\fN+1}-C_{2\fN+1}$};
\node[draw=none] (g4) at (6,0) {$D_1$};
\node[draw=none] (g5) at (4.5,1) {$D_1$};
\node[draw=none] (g6) at (4.5,-1) {$D_1$};
\node[draw=none] (f2) at (4.5,2) {$[\fN-1]_2$};
\node[draw=none] (f3) at (7.5,0) {$[\fN-1]_2$};
\draw[black] (gm1)--(g0)--(g1);
\draw[thick,red](g1)--(g4);
\draw[thick,red](g1) -- node[midway,above left,xshift=0.2cm] {\scriptsize $2\fN-1$}  (g5);
\draw[thick](g1)--(g6);
\draw[thick,gray](g5)--(g6);
\draw[thick,gray](g6)-- node[below right,midway]{\scriptsize $1$} (g4);
\draw[thick,blue](g5) --node[above right,midway,xshift=-0.4cm]{\scriptsize$2\fN-1$} (g4);
\draw[thick,draw,solid,black,snake it] (g5) --(f2);
\draw[thick,draw,solid,black,snake it] (g4) -- (f3);
\end{tikzpicture}
}
where the tail $D_1-C_1-\cdots-D_{2\fN+1}-C_{2\fN+1}$ gives rise to the $\SO(4\fN+4)$ CB symmetry corresponding to the part $1^{4\fN+4}$ of the partition, whereas the balanced gauge group $C_{3\fN}$ gives rise to the $\SO(2)$ CB symmetry corresponding to the part $(2\fN-3)^2$ of the partition.  Note that the $D_{2\fN+2}$ gauge node is overbalanced.  Indeed, the tail on the left is determined by the $T_{[(2\fN-3)^2, 1^{4\fN+4}]}(\SO(8\fN-2))$ theory.

On the other hand, the mirror theory for $D_{4\fN}(\SO(4\fN+4))$ is described by \eref{mirr4NSO4Np4a}: 
\bes{ \label{mirr4NSO4Np4}
\begin{tikzpicture}[baseline]
\node[draw=none] (D10) at (6.5,0) {$C_{2\fN}-D_{2\fN}-\cdots-C_1-D_1$};
\node[draw=none] (D2N) at (2.5,0) {$D_{2 \fN+1}$};
\node[draw=none] (C1) at (0.8,0) {$C_{2\fN}$};
\node[draw=none] (F) at (-3,1) {$[\fN-1]_2$};
\node[draw=none] (D1a) at (-1,1) {$D_1$};
\node[draw=none] (D1b) at (-1,-1){$D_1$};
\draw[solid, thick, snake it] (F) to (D1a);
\draw[solid] (C1)--(D2N)--(D10);
\draw[solid,red,thick] (C1) -- node[midway,below left] {\scriptsize $2\fN-1$} (D1a) ;
\draw[solid, very thick] (C1) to (D1b);
\draw[solid, thick, gray] (D2N) -- node[midway,above right] {\scriptsize $1$} (D1a);
\end{tikzpicture}
}

Now we glue \eref{mirr4NSO4Np4} and \eref{mirrD8NSO8Nm2} together by fusing the tails $D_1-C_1-\cdots-D_{2\fN+1}-C_{2\fN+1}$ of the two quivers. This corresponds to commonly gauging the CB symmetry $\SO(4\fN+4)$ of the two theories, as instructed by \eref{cusp2a}.  However, there is no explicit $C_{2\fN+1}$ gauge group in \eref{mirr4NSO4Np4}, but the $D_{2\fN+1}$ node is connected to the $C_{2\fN}$ node and a $D_1$ node (connected to the wiggle line).  In other word, the aforementioned $C_{2\fN+1}$ group should be viewed as split into $C_{2\fN}$ and $C_1$, where the $D_1$ subgroup of the latter is gauged. Thus, upon gluing the two theories, we also need to split the $C_{3\fN}$ node of \eref{mirrD8NSO8Nm2} into $C_{3\fN-1}$ and $C_1$, where the latter $C_1$ is then identified with the $C_1$ group mentioned earlier whose $D_1$ subgroup is gauged.  The central node thus becomes $C_{3\fN-1}$ as in \eref{eq:6massesDnDmwithm=1}. The mirror theory in question thus acquires the left part of \eref{mirr4NSO4Np4}.  The lower left $D_1$ node in \eref{mirr4NSO4Np4} becomes the lower left $D_1$ node in \eref{eq:6massesDnDmwithm=1}. This part of gluing indeed explains the $D_1-C_{3\fn-1}$ tail in \eref{mirr6masses}.  Moreover, each $D_1$ node connected to the wiggle lines are connected together as in the previous case of gluing.  We thus arrive at \eref{eq:6massesDnDmwithm=1}, as required.

\acknowledgments

We thank C. Closset for useful discussions. F.C. is supported by STFC consolidated grant ST/T000708/1. The work of S.G. is supported in part by the ERC Consolidator Grant number 682608 ``Higgs bundles: Supersymmetric Gauge Theories and Geometry (HIGGSBNDL)" and partly by the INFN grant ``Per attività di formazione per sostenere progetti di ricerca'' (GRANT 73/STRONGQFT).. N. M. thanks Stefano Lionetti for his hospitality at \href{https://www.zetalab.com/}{Zetalab} during the completion of the project. A. M. received funding from ``la Caixa" Foundation (ID 100010434) with fellowship code LCF/BQ/IN18/11660045 and from the European Union’s Horizon 2020 research and innovation programme under the Marie Sk\l odowska-Curie grant agreement No. 713673 until September 2021. The work of A. M. is supported in part by Deutsche Forschungsgemeinschaft under Germany's Excellence Strategy EXC 2121 Quantum Universe 390833306.

\appendix

\section{Some properties of $D_p(\USp(2N))$}\label{appendixusp}

In this appendix, we collect some properties of the $D_p(\USp(2N))$ that we have found.  This theory was defined and discussed in \cite[(2.11)--(2.14)]{Carta:2021whq}.

First, similarly to \cref{eq:DnDmDecompCase1,eq:DnDmDecompCase2,eq:DnDmDecompCase3,eq:DnDmDecompCase4,eq:DnDmDecompCase5}, we find that a certain subclass of the $D_p(\USp(2N))$ theories admits a weakly coupling cusp in the conformal manifold:
\begin{equation}
\begin{split}
     D_{2\mu+2}&(\USp((2 \mu +2) (2 \mathfrak{m}-1)))=\\
     =&\,D_{2\mu}(\USp(2 \mu  (2 \mathfrak{m}-1)))-\USp(2 \mu  (2 \mathfrak{m}-1))-D_2(\USp((4 \mu +2) (2 \mathfrak{m}-1)))\fstop
\end{split}
\end{equation}
where $\mu \geq 1$ and $\fm \geq 1$.

Secondly, we find that whenever $N$ is a multiple of $p$, the $D_p(\USp(2N))$ theory admits a Lagrangian description.  Let us analyze two cases according to the parity of $p$.

\subsubsection*{The case of $p$ even} 
In this case, we write
\begin{equation}
    p=2\mu~, \quad N= \fm p = 2 \fm \mu~, \qquad \mu,\, \fm\geq 1\fstop
\end{equation}
The Lagrangian description of $D_{2\mu}(\USp(4\mathfrak{m}\mu))$ can be written as
\begin{equation} \label{lagrangianA}
    D_{\fm+1}-C_{2\fm}-D_{3\fm+1}-\cdots- D_{(2\mu-1)\fm+1}-[C_{2\mu\fm}]\fstop
\end{equation}
As we commented around Footnote \ref{caution}, it is not clear whether reduction of this theory on a circle to 3d yields a 3d $\CN=4$ gauge theory with the same quiver description. Nevertheless, if we view \eref{lagrangianA} as a $3$d $\mathcal{N}=4$ gauge theory, this is a quiver description of the  $T^{\sigma}_{\rho}[\USp(4\mathfrak{m}\mu)]$ theory, with 
\begin{equation}
    \sigma = \left[1^{4\mu \mathfrak{m}}\right] \aand \rho =\left[(2\mathfrak{m})^{2\mu},1\right]~,
\end{equation}
whose mirror theory is $T^{\rho}_{\sigma}[\SO(4\mathfrak{m}\mu+1)]$ and can be described by

\begin{equation} \label{mirrA}
\scalebox{0.75}{
$
\begin{array}{lll}
&C_\mu-B_{2\mu+1}-\cdots-B_{(2\mu-1)\mathfrak{m}-3}-C_{(2\mu-1)\mathfrak{m}-1}-&B_{(2\mu-1)\mathfrak{m}}-C_{(2\mu-1)\mathfrak{m}N}-B_{(2\mu-1)\mathfrak{m}N-1}-C_{(2\mu-1)\mathfrak{m}-1}- \cdots -B_{1}-C_{1}-B_{0} \\
&\,\, | &\,\, | \\
&\!\![B_{0}] &\![C_{\mu}]
\end{array}$
}
\end{equation}

We observe that the Higgs branch dimension of the mirror theory \eref{defsigma} is exactly equal to the rank of the corresponding 4d theory. However, the Coulomb branch dimension of the mirror theory is larger than the value $24(c-a)$ of the corresponding 4d theory. As we commented in Footnote \ref{caution}, it is not clear whether $24(c-a)$ is equal to the Higgs branch of the 4d theory, since the orthogonal gauge groups may not be completely Higgsed at a generic point on the Higgs branch. This is due to the fact that the orthogonal gauge group which is conformal in 4d is underbalanced in 3d; see an explicit example in \eref{DSQCD} below.

As an example, for $\mu=1$, the 4d theory is simply an $\SO(2\fm+2)$ SQCD with $2\fm$ flavors:
\begin{equation} \label{DSQCD}
\begin{split}
     D_2(\USp(4\fm))& \,:\; D_{\fm+1}-[C_{2\fm}]\fstop
\end{split}
\end{equation}
Viewing this as a 3d $\CN=4$ gauge theory, the mirror theory is given by \cite[Figure 17]{Feng:2000eq}
\bes{
\scalebox{0.9}{
$\begin{array}{lll}
&C_1-B_1-\cdots-B_{\fm-1}-C_\fm-&B_{\fm}-C_{\fm}-B_{\fm-1}-C_{\fm-1}- \cdots -B_{1}-C_{1}-B_0 \\
&\,\, | &\,\, | \\
&\![B_0] &\![C_1]
\end{array}$
}
}

\subsubsection*{The case of $p$ odd}

In this case, we write
\begin{equation}
    p=2\mu+1~, \quad N= \fm p = \fm (2\mu+1)~, \qquad \mu,\, \fm\geq 1\fstop
\end{equation}
The Lagrangian description of $D_{2\mu+1}(\USp(2(2\mu+1)\mathfrak{m}))$ can be written as
\begin{equation}
    [D_2]-C_{\fm}-D_{2\fm+1}-C_{3\fm}-\cdots - D_{2\mu\fm +1} - [C_{(2\mu+1)\fm}]\fstop
\end{equation}
Viewing it a $3$d $\mathcal{N}=4$ gauge theory, this is in fact a quiver description of the  $T^{\sigma}_{\rho}\left[\USp( (4\mu+2)\mathfrak{m}+8\mu)\right]$ theory, with
\begin{equation}
    \sigma = \left[(2\mu)^{4},1^{2(2\mu+1)\mathfrak{m}}\right] \aand \rho =\left[(2\mathfrak{m}+4)^{2\mu},2\mathfrak{m}+1\right]\fstop
\end{equation}
whose mirror theory is $T^{\rho}_{\sigma}[\SO( (4\mu+2)\mathfrak{m}+8\mu+1)]$.  The same comments below \eref{mirrA} apply here.

As an example, for $\mu=1$, we have
\begin{equation}
    \begin{split}
        D_3(\USp(6\fm))&\,:\; [D_2]-C_{\fm}-D_{2\fm+1}-[C_{3\fm}]\fstop
    \end{split}
\end{equation}
Viewing this as a 3d $\CN=4$ gauge theory, the mirror theory can be described as
\bes{
\scalebox{0.8}{
$\begin{array}{lll}
D_1-C_1-D_2-C_3-D_4-C_5-\cdots-D_{2\fm-2}-&C_{2\fm-1}-B_{2\fm-1}-C_{2\fm}-&B_{2\fm}-C_{2\fm}-B_{2\fm-1}-\cdots - C_1-B_0 \\
&| &| \\
&\!\![B_0] &\!\![C_1]
\end{array}
$
}
}

\bibliographystyle{JHEP}
\bibliography{mybib}

\providecommand{\href}[2]{#2}\begingroup\raggedright\begin{thebibliography}{10}

\bibitem{Seiberg:1994rs}
N.~Seiberg and E.~Witten, \emph{{Electric - magnetic duality, monopole
  condensation, and confinement in N=2 supersymmetric Yang-Mills theory}},
  \href{https://doi.org/10.1016/0550-3213(94)90124-4}{\emph{Nucl. Phys. B}
  {\bfseries 426} (1994) 19}
  [\href{https://arxiv.org/abs/hep-th/9407087}{{\ttfamily hep-th/9407087}}].

\bibitem{Seiberg:1994aj}
N.~Seiberg and E.~Witten, \emph{{Monopoles, duality and chiral symmetry
  breaking in N=2 supersymmetric QCD}},
  \href{https://doi.org/10.1016/0550-3213(94)90214-3}{\emph{Nucl. Phys.}
  {\bfseries B431} (1994) 484}
  [\href{https://arxiv.org/abs/hep-th/9408099}{{\ttfamily hep-th/9408099}}].

\bibitem{Argyres:1995jj}
P.~C. Argyres and M.~R. Douglas, \emph{{New phenomena in SU(3) supersymmetric
  gauge theory}},
  \href{https://doi.org/10.1016/0550-3213(95)00281-V}{\emph{Nucl. Phys. B}
  {\bfseries 448} (1995) 93}
  [\href{https://arxiv.org/abs/hep-th/9505062}{{\ttfamily hep-th/9505062}}].

\bibitem{Argyres:1995xn}
P.~C. Argyres, M.~Plesser, N.~Seiberg and E.~Witten, \emph{{New N=2
  superconformal field theories in four-dimensions}},
  \href{https://doi.org/10.1016/0550-3213(95)00671-0}{\emph{Nucl. Phys. B}
  {\bfseries 461} (1996) 71}
  [\href{https://arxiv.org/abs/hep-th/9511154}{{\ttfamily hep-th/9511154}}].

\bibitem{Eguchi:1996vu}
T.~Eguchi, K.~Hori, K.~Ito and S.-K. Yang, \emph{{Study of N=2 superconformal
  field theories in four-dimensions}},
  \href{https://doi.org/10.1016/0550-3213(96)00188-5}{\emph{Nucl. Phys. B}
  {\bfseries 471} (1996) 430}
  [\href{https://arxiv.org/abs/hep-th/9603002}{{\ttfamily hep-th/9603002}}].

\bibitem{Eguchi:1996ds}
T.~Eguchi and K.~Hori, \emph{{N=2 superconformal field theories in
  four-dimensions and A-D-E classification}},  in \emph{{Conference on the
  Mathematical Beauty of Physics (In Memory of C. Itzykson)}}, pp.~67--82, 7,
  1996, \href{https://arxiv.org/abs/hep-th/9607125}{{\ttfamily
  hep-th/9607125}}.

\bibitem{Gaiotto:2009we}
D.~Gaiotto, \emph{{N=2 dualities}},
  \href{https://doi.org/10.1007/JHEP08(2012)034}{\emph{JHEP} {\bfseries 08}
  (2012) 034} [\href{https://arxiv.org/abs/0904.2715}{{\ttfamily 0904.2715}}].

\bibitem{Gaiotto:2009hg}
D.~Gaiotto, G.~W. Moore and A.~Neitzke, \emph{{Wall-crossing, Hitchin Systems,
  and the WKB Approximation}},
  \href{https://arxiv.org/abs/0907.3987}{{\ttfamily 0907.3987}}.

\bibitem{Ohmori:2015pua}
K.~Ohmori, H.~Shimizu, Y.~Tachikawa and K.~Yonekura, \emph{{6d
  $\mathcal{N}=(1,0)$ theories on $T^2$ and class S theories: Part I}},
  \href{https://doi.org/10.1007/JHEP07(2015)014}{\emph{JHEP} {\bfseries 07}
  (2015) 014} [\href{https://arxiv.org/abs/1503.06217}{{\ttfamily
  1503.06217}}].

\bibitem{Ohmori:2015pia}
K.~Ohmori, H.~Shimizu, Y.~Tachikawa and K.~Yonekura, \emph{{6d
  $\mathcal{N}{=}(1,0)$ theories on $S^1/T^2$ and class S theories: part II}},
  \href{https://arxiv.org/abs/1508.00915}{{\ttfamily 1508.00915}}.

\bibitem{Ohmori:2018ona}
K.~Ohmori, Y.~Tachikawa and G.~Zafrir, \emph{{Compactifications of 6d $N = (1,
  0)$ SCFTs with non-trivial Stiefel-Whitney classes}},
  \href{https://doi.org/10.1007/JHEP04(2019)006}{\emph{JHEP} {\bfseries 04}
  (2019) 006} [\href{https://arxiv.org/abs/1812.04637}{{\ttfamily
  1812.04637}}].

\bibitem{Katz:1996fh}
S.~H. Katz, A.~Klemm and C.~Vafa, \emph{{Geometric engineering of quantum field
  theories}}, \href{https://doi.org/10.1016/S0550-3213(97)00282-4}{\emph{Nucl.
  Phys. B} {\bfseries 497} (1997) 173}
  [\href{https://arxiv.org/abs/hep-th/9609239}{{\ttfamily hep-th/9609239}}].

\bibitem{Shapere:1999xr}
A.~D. Shapere and C.~Vafa, \emph{{BPS structure of Argyres-Douglas
  superconformal theories}},
  \href{https://arxiv.org/abs/hep-th/9910182}{{\ttfamily hep-th/9910182}}.

\bibitem{Intriligator:1996ex}
K.~A. Intriligator and N.~Seiberg, \emph{{Mirror symmetry in three-dimensional
  gauge theories}},
  \href{https://doi.org/10.1016/0370-2693(96)01088-X}{\emph{Phys. Lett.}
  {\bfseries B387} (1996) 513}
  [\href{https://arxiv.org/abs/hep-th/9607207}{{\ttfamily hep-th/9607207}}].

\bibitem{Ferlito:2017xdq}
G.~Ferlito, A.~Hanany, N.~Mekareeya and G.~Zafrir, \emph{{3d Coulomb branch and
  5d Higgs branch at infinite coupling}},
  \href{https://doi.org/10.1007/JHEP07(2018)061}{\emph{JHEP} {\bfseries 07}
  (2018) 061} [\href{https://arxiv.org/abs/1712.06604}{{\ttfamily
  1712.06604}}].

\bibitem{Cabrera:2018jxt}
S.~Cabrera, A.~Hanany and F.~Yagi, \emph{{Tropical Geometry and Five
  Dimensional Higgs Branches at Infinite Coupling}},
  \href{https://doi.org/10.1007/JHEP01(2019)068}{\emph{JHEP} {\bfseries 01}
  (2019) 068} [\href{https://arxiv.org/abs/1810.01379}{{\ttfamily
  1810.01379}}].

\bibitem{Hanany:2018uhm}
A.~Hanany and N.~Mekareeya, \emph{{The small E$_{8}$ instanton and the Kraft
  Procesi transition}},
  \href{https://doi.org/10.1007/JHEP07(2018)098}{\emph{JHEP} {\bfseries 07}
  (2018) 098} [\href{https://arxiv.org/abs/1801.01129}{{\ttfamily
  1801.01129}}].

\bibitem{Cabrera:2019izd}
S.~Cabrera, A.~Hanany and M.~Sperling, \emph{{Magnetic quivers, Higgs branches,
  and 6d $N$=(1,0) theories}},
  \href{https://doi.org/10.1007/JHEP06(2019)071}{\emph{JHEP} {\bfseries 06}
  (2019) 071} [\href{https://arxiv.org/abs/1904.12293}{{\ttfamily
  1904.12293}}].

\bibitem{Akhond:2020vhc}
M.~Akhond, F.~Carta, S.~Dwivedi, H.~Hayashi, S.-S. Kim and F.~Yagi,
  \emph{{Five-brane webs, Higgs branches and unitary/orthosymplectic magnetic
  quivers}}, \href{https://doi.org/10.1007/JHEP12(2020)164}{\emph{JHEP}
  {\bfseries 12} (2020) 164}
  [\href{https://arxiv.org/abs/2008.01027}{{\ttfamily 2008.01027}}].

\bibitem{Akhond:2021knl}
M.~Akhond, F.~Carta, S.~Dwivedi, H.~Hayashi, S.-S. Kim and F.~Yagi,
  \emph{{Factorised 3d $\mathcal{N}=4$ orthosymplectic quivers}},
  \href{https://doi.org/10.1007/JHEP05(2021)269}{\emph{JHEP} {\bfseries 05}
  (2021) 269} [\href{https://arxiv.org/abs/2101.12235}{{\ttfamily
  2101.12235}}].

\bibitem{Akhond:2021ffo}
M.~Akhond and F.~Carta, \emph{{Magnetic quivers from brane webs with
  O7$^+$-planes}},  \href{https://arxiv.org/abs/2107.09077}{{\ttfamily
  2107.09077}}.

\bibitem{Bourget:2020gzi}
A.~Bourget, J.~F. Grimminger, A.~Hanany, M.~Sperling and Z.~Zhong,
  \emph{{Magnetic Quivers from Brane Webs with O5 Planes}},
  \href{https://doi.org/10.1007/JHEP07(2020)204}{\emph{JHEP} {\bfseries 07}
  (2020) 204} [\href{https://arxiv.org/abs/2004.04082}{{\ttfamily
  2004.04082}}].

\bibitem{Bourget:2020asf}
A.~Bourget, J.~F. Grimminger, A.~Hanany, M.~Sperling, G.~Zafrir and Z.~Zhong,
  \emph{{Magnetic quivers for rank 1 theories}},
  \href{https://doi.org/10.1007/JHEP09(2020)189}{\emph{JHEP} {\bfseries 09}
  (2020) 189} [\href{https://arxiv.org/abs/2006.16994}{{\ttfamily
  2006.16994}}].

\bibitem{Bourget:2020mez}
A.~Bourget, S.~Giacomelli, J.~F. Grimminger, A.~Hanany, M.~Sperling and
  Z.~Zhong, \emph{{S-fold magnetic quivers}},
  \href{https://doi.org/10.1007/JHEP02(2021)054}{\emph{JHEP} {\bfseries 02}
  (2021) 054} [\href{https://arxiv.org/abs/2010.05889}{{\ttfamily
  2010.05889}}].

\bibitem{boalch2008irregular}
P.~Boalch, \emph{{Irregular connections and Kac-Moody root systems}},
  \href{https://arxiv.org/abs/0806.1050}{{\ttfamily 0806.1050}}.

\bibitem{Xie:2012hs}
D.~Xie, \emph{{General Argyres-Douglas Theory}},
  \href{https://doi.org/10.1007/JHEP01(2013)100}{\emph{JHEP} {\bfseries 01}
  (2013) 100} [\href{https://arxiv.org/abs/1204.2270}{{\ttfamily 1204.2270}}].

\bibitem{Xie:2013jc}
D.~Xie and P.~Zhao, \emph{{Central charges and RG flow of strongly-coupled N=2
  theory}}, \href{https://doi.org/10.1007/JHEP03(2013)006}{\emph{JHEP}
  {\bfseries 03} (2013) 006} [\href{https://arxiv.org/abs/1301.0210}{{\ttfamily
  1301.0210}}].

\bibitem{DelZotto:2014kka}
M.~Del~Zotto and A.~Hanany, \emph{{Complete Graphs, Hilbert Series, and the
  Higgs branch of the 4d $\mathcal{N} =$ 2 $(A_n,A_m)$ SCFTs}},
  \href{https://doi.org/10.1016/j.nuclphysb.2015.03.017}{\emph{Nucl. Phys. B}
  {\bfseries 894} (2015) 439}
  [\href{https://arxiv.org/abs/1403.6523}{{\ttfamily 1403.6523}}].

\bibitem{Buican:2015hsa}
M.~Buican and T.~Nishinaka, \emph{{Argyres\textendash{}Douglas theories, S$^1$
  reductions, and topological symmetries}},
  \href{https://doi.org/10.1088/1751-8113/49/4/045401}{\emph{J. Phys. A}
  {\bfseries 49} (2016) 045401}
  [\href{https://arxiv.org/abs/1505.06205}{{\ttfamily 1505.06205}}].

\bibitem{Xie:2017vaf}
D.~Xie and S.-T. Yau, \emph{{Argyres-Douglas matter and N=2 dualities}},
  \href{https://arxiv.org/abs/1701.01123}{{\ttfamily 1701.01123}}.

\bibitem{Benvenuti:2017kud}
S.~Benvenuti and S.~Giacomelli, \emph{{Abelianization and sequential
  confinement in $2+1$ dimensions}},
  \href{https://doi.org/10.1007/JHEP10(2017)173}{\emph{JHEP} {\bfseries 10}
  (2017) 173} [\href{https://arxiv.org/abs/1706.04949}{{\ttfamily
  1706.04949}}].

\bibitem{Benvenuti:2017bpg}
S.~Benvenuti and S.~Giacomelli, \emph{{Lagrangians for generalized
  Argyres-Douglas theories}},
  \href{https://doi.org/10.1007/JHEP10(2017)106}{\emph{JHEP} {\bfseries 10}
  (2017) 106} [\href{https://arxiv.org/abs/1707.05113}{{\ttfamily
  1707.05113}}].

\bibitem{Dey:2020hfe}
A.~Dey, \emph{{Three Dimensional Mirror Symmetry beyond $ADE$ quivers and
  Argyres-Douglas theories}},
  \href{https://arxiv.org/abs/2004.09738}{{\ttfamily 2004.09738}}.

\bibitem{Closset:2020scj}
C.~Closset, S.~Schafer-Nameki and Y.-N. Wang, \emph{{Coulomb and Higgs Branches
  from Canonical Singularities: Part 0}},
  \href{https://doi.org/10.1007/JHEP02(2021)003}{\emph{JHEP} {\bfseries 02}
  (2021) 003} [\href{https://arxiv.org/abs/2007.15600}{{\ttfamily
  2007.15600}}].

\bibitem{Closset:2020afy}
C.~Closset, S.~Giacomelli, S.~Schafer-Nameki and Y.-N. Wang, \emph{{5d and 4d
  SCFTs: Canonical Singularities, Trinions and S-Dualities}},
  \href{https://arxiv.org/abs/2012.12827}{{\ttfamily 2012.12827}}.

\bibitem{Giacomelli:2020ryy}
S.~Giacomelli, N.~Mekareeya and M.~Sacchi, \emph{{New aspects of
  Argyres--Douglas theories and their dimensional reduction}},
  \href{https://doi.org/10.1007/JHEP03(2021)242}{\emph{JHEP} {\bfseries 03}
  (2021) 242} [\href{https://arxiv.org/abs/2012.12852}{{\ttfamily
  2012.12852}}].

\bibitem{Carta:2021whq}
F.~Carta, S.~Giacomelli, N.~Mekareeya and A.~Mininno, \emph{{Conformal
  manifolds and 3d mirrors of Argyres-Douglas theories}},
  \href{https://doi.org/10.1007/JHEP08(2021)015}{\emph{JHEP} {\bfseries 08}
  (2021) 015} [\href{https://arxiv.org/abs/2105.08064}{{\ttfamily
  2105.08064}}].

\bibitem{Xie:2021ewm}
D.~Xie, \emph{{3d mirror for Argyres-Douglas theories}},
  \href{https://arxiv.org/abs/2107.05258}{{\ttfamily 2107.05258}}.

\bibitem{Dey:2021rxw}
A.~Dey, \emph{{Higgs Branches of Argyres-Douglas theories as Quiver
  Varieties}},  \href{https://arxiv.org/abs/2109.07493}{{\ttfamily
  2109.07493}}.

\bibitem{Cecotti:2010fi}
S.~Cecotti, A.~Neitzke and C.~Vafa, \emph{{R-Twisting and 4d/2d
  Correspondences}},  \href{https://arxiv.org/abs/1006.3435}{{\ttfamily
  1006.3435}}.

\bibitem{Kapustin:1998fa}
A.~Kapustin, \emph{{D(n) quivers from branes}},
  \href{https://doi.org/10.1088/1126-6708/1998/12/015}{\emph{JHEP} {\bfseries
  12} (1998) 015} [\href{https://arxiv.org/abs/hep-th/9806238}{{\ttfamily
  hep-th/9806238}}].

\bibitem{Gaiotto:2008ak}
D.~Gaiotto and E.~Witten, \emph{{S-Duality of Boundary Conditions In N=4 Super
  Yang-Mills Theory}},
  \href{https://doi.org/10.4310/ATMP.2009.v13.n3.a5}{\emph{Adv. Theor. Math.
  Phys.} {\bfseries 13} (2009) 721}
  [\href{https://arxiv.org/abs/0807.3720}{{\ttfamily 0807.3720}}].

\bibitem{Buican:2014hfa}
M.~Buican, S.~Giacomelli, T.~Nishinaka and C.~Papageorgakis,
  \emph{{Argyres-Douglas Theories and S-Duality}},
  \href{https://doi.org/10.1007/JHEP02(2015)185}{\emph{JHEP} {\bfseries 02}
  (2015) 185} [\href{https://arxiv.org/abs/1411.6026}{{\ttfamily 1411.6026}}].

\bibitem{Buican:2017fiq}
M.~Buican, Z.~Laczko and T.~Nishinaka, \emph{{$ \mathcal{N} $ = 2 S-duality
  revisited}}, \href{https://doi.org/10.1007/JHEP09(2017)087}{\emph{JHEP}
  {\bfseries 09} (2017) 087}
  [\href{https://arxiv.org/abs/1706.03797}{{\ttfamily 1706.03797}}].

\bibitem{Buican:2021xhs}
M.~Buican and H.~Jiang, \emph{{1-Form Symmetry, Isolated N=2 SCFTs, and
  Calabi-Yau Threefolds}},  \href{https://arxiv.org/abs/2106.09807}{{\ttfamily
  2106.09807}}.

\bibitem{Cecotti:2012jx}
S.~Cecotti and M.~Del~Zotto, \emph{{Infinitely many N=2 SCFT with ADE flavor
  symmetry}}, \href{https://doi.org/10.1007/JHEP01(2013)191}{\emph{JHEP}
  {\bfseries 01} (2013) 191} [\href{https://arxiv.org/abs/1210.2886}{{\ttfamily
  1210.2886}}].

\bibitem{Cecotti:2013lda}
S.~Cecotti, M.~Del~Zotto and S.~Giacomelli, \emph{{More on the N=2
  superconformal systems of type $D_p(G)$}},
  \href{https://doi.org/10.1007/JHEP04(2013)153}{\emph{JHEP} {\bfseries 04}
  (2013) 153} [\href{https://arxiv.org/abs/1303.3149}{{\ttfamily 1303.3149}}].

\bibitem{Wang:2015mra}
Y.~Wang and D.~Xie, \emph{{Classification of Argyres-Douglas theories from M5
  branes}}, \href{https://doi.org/10.1103/PhysRevD.94.065012}{\emph{Phys. Rev.
  D} {\bfseries 94} (2016) 065012}
  [\href{https://arxiv.org/abs/1509.00847}{{\ttfamily 1509.00847}}].

\bibitem{Wang:2018gvb}
Y.~Wang and D.~Xie, \emph{{Codimension-two defects and Argyres-Douglas theories
  from outer-automorphism twist in 6d $(2,0)$ theories}},
  \href{https://doi.org/10.1103/PhysRevD.100.025001}{\emph{Phys. Rev. D}
  {\bfseries 100} (2019) 025001}
  [\href{https://arxiv.org/abs/1805.08839}{{\ttfamily 1805.08839}}].

\bibitem{Nanopoulos:2010bv}
D.~Nanopoulos and D.~Xie, \emph{{More Three Dimensional Mirror Pairs}},
  \href{https://doi.org/10.1007/JHEP05(2011)071}{\emph{JHEP} {\bfseries 05}
  (2011) 071} [\href{https://arxiv.org/abs/1011.1911}{{\ttfamily 1011.1911}}].

\bibitem{Brandhuber:1995zp}
A.~Brandhuber and K.~Landsteiner, \emph{{On the monodromies of N=2
  supersymmetric Yang-Mills theory with gauge group SO(2n)}},
  \href{https://doi.org/10.1016/0370-2693(95)00986-U}{\emph{Phys. Lett. B}
  {\bfseries 358} (1995) 73}
  [\href{https://arxiv.org/abs/hep-th/9507008}{{\ttfamily hep-th/9507008}}].

\bibitem{Caibar:1999aaa}
M.~Caibar, \emph{{Minimal models of canonical singularities and their
  cohomology}}, Ph.D. thesis, University of Warwick, Warwick, U.K, 1999.

\bibitem{Caibar2003:aaa}
M.~Caibar, \emph{On the number of crepant valuations of canonical
  singularities},
  \href{https://doi.org/https://doi.org/10.1112/S0024610703004514}{\emph{Journal
  of the London Mathematical Society} {\bfseries 68} (2003) 307}.

\bibitem{Sagemath}
{The Sage Developers}, \emph{{S}ageMath, the {S}age {M}athematics {S}oftware
  {S}ystem ({V}ersion 9.3)}, 2021.

\bibitem{Gaiotto:2012uq}
D.~Gaiotto and S.~S. Razamat, \emph{{Exceptional Indices}},
  \href{https://doi.org/10.1007/JHEP05(2012)145}{\emph{JHEP} {\bfseries 05}
  (2012) 145} [\href{https://arxiv.org/abs/1203.5517}{{\ttfamily 1203.5517}}].

\bibitem{Feng:2000eq}
B.~Feng and A.~Hanany, \emph{{Mirror symmetry by O3 planes}},
  \href{https://doi.org/10.1088/1126-6708/2000/11/033}{\emph{JHEP} {\bfseries
  11} (2000) 033} [\href{https://arxiv.org/abs/hep-th/0004092}{{\ttfamily
  hep-th/0004092}}].

\bibitem{Carta:2020plx}
F.~Carta and A.~Mininno, \emph{{No go for a flow}},
  \href{https://doi.org/10.1007/JHEP05(2020)108}{\emph{JHEP} {\bfseries 05}
  (2020) 108} [\href{https://arxiv.org/abs/2002.07816}{{\ttfamily
  2002.07816}}].

\bibitem{Benini:2010uu}
F.~Benini, Y.~Tachikawa and D.~Xie, \emph{{Mirrors of 3d Sicilian theories}},
  \href{https://doi.org/10.1007/JHEP09(2010)063}{\emph{JHEP} {\bfseries 09}
  (2010) 063} [\href{https://arxiv.org/abs/1007.0992}{{\ttfamily 1007.0992}}].

\bibitem{Beratto:2020wmn}
E.~Beratto, S.~Giacomelli, N.~Mekareeya and M.~Sacchi, \emph{{3d mirrors of the
  circle reduction of twisted A$_{2N}$ theories of class S}},
  \href{https://doi.org/10.1007/JHEP09(2020)161}{\emph{JHEP} {\bfseries 09}
  (2020) 161} [\href{https://arxiv.org/abs/2007.05019}{{\ttfamily
  2007.05019}}].

\end{thebibliography}\endgroup

\end{document}